\documentclass[sn-mathphys, Numbered]{sn-jnl}% Math and Physical Sciences Reference Style
\usepackage{graphicx}%
\usepackage{multirow}%
\usepackage{amsmath,amssymb,amsfonts}%
\usepackage{amsthm}%
\usepackage{mathrsfs}%
\usepackage[title]{appendix}%
\usepackage{xcolor}%
\usepackage{textcomp}%
\usepackage{manyfoot}%
\usepackage{booktabs}%
\usepackage{algorithm}%
\usepackage{algorithmicx}%
\usepackage{algpseudocode}%
\usepackage{listings}%
\usepackage{bbm}
\newcommand{\RR}{\mathbb{R}}
\newcommand{\ZZ}{\mathbb{Z}}
\newcommand{\PP}{\mathbb{P}}
\newcommand{\EE}{\mathbb{E}}
\newcommand{\TT}{\mathbb{T}}
\newcommand{\NN}{\mathbb{N}}
\renewcommand{\SS}{\mathbb{S}}

\newcommand{\x}{\boldsymbol{x}}{}
\newcommand{\y}{\boldsymbol{y}}{}
\newcommand{\e}{\boldsymbol{e}}{}
\newcommand{\z}{\boldsymbol{z}}{}
\renewcommand{\u}{\boldsymbol{u}}{}

\newcommand{\tw}{\tilde}{}
\newcommand{\p}{\partial}{}
\renewcommand{\div  }{\nabla \cdot}
\newcommand{\pt  }{\p_t}
\newcommand{\dd}{\mathrm{d}}{}

\newcommand{\id}{\mathbbm{1}}{}
\newcommand{\ep}{\varepsilon}{}

\newcommand{\beq}{\begin{equation}}
\newcommand{\eeq}{\end{equation}}

\definecolor{purple}{rgb}{0.6, 0.0, 1.0}

\definecolor{mygreen}{rgb}{0.0,0.55,0.3}

\newcommand{\model}{Active Lattice Gas}
\newcommand{\macc}{ALG}

\renewcommand{\Re}[1]{\operatorname{Re}({#1})}

\newtheorem{theorem}{\bf Theorem}[section]

\newtheorem{proposition}{\bf Proposition}[section]

\newtheorem{definition}{\bf Definition}[section]
\newtheorem{remark}{\bf Remark}[section]
\PassOptionsToPackage{unicode}{hyperref}
\PassOptionsToPackage{naturalnames}{hyperref}
\begin{document}

\title[Hydrodynamics and phase separation for an active exclusion process]{Exact hydrodynamics and onset of phase separation for an active exclusion process}

% GivenName	-> \fnm{Joergen W.}

\author*{\fnm{James} \sur{Mason}$^1$}\email{jm2386@cam.ac.uk}

\author{\fnm{Cl\'ement} \sur{Erignoux}$^{2}$}

\author{\fnm{Robert L} \sur{Jack}$^{1,3}$}%\email{rlj22@cam.ac.uk}

\author{\fnm{Maria} \sur{Bruna}$^1$}%\email{bruna@maths.cam.ac.uk}

\affil{$^1$ \orgdiv{Department of Applied Mathematics and Theoretical Physics}, \orgname{University of Cambridge}, \orgaddress{\street{Wilberforce Road}, \city{Cambridge}, \postcode{CB3 0WA}, \country{UK}}}

\affil{$^2$ \orgdiv{INRIA, Laboratoire Paul Painlev\'e}, \orgname{University of Lille, CNRS, UMR 8524}, \orgaddress{\street{40 Avenue Halley Bâtiment B}, \city{Lille}, \postcode{59650 Villeneuve-d'Ascq}, \country{France}}}

\affil{$^3$ \orgdiv{Yusuf Hamied Department of Chemistry}, \orgname{University of Cambridge}, \orgaddress{\street{Lensfield Road}, \city{Cambridge}, \postcode{CB2 1EW}, \country{UK}}}


\abstract{
We consider a lattice model of active matter with exclusion and derive its hydrodynamic description exactly. The hydrodynamic limit leads to an integro-differential equation for the density of particles with a given orientation. Volume exclusion results in nonlinear mobility dependent on spatial density. Such models of active matter can support motility-induced phase separation, which occurs despite the absence of attractive interactions. We study the onset of phase separation with linear stability analysis and numerical simulations.
}




\maketitle

%%%%%%%%%%%%%%%%%%%%%%%%%%%
\newcommand{  \pas}{f^p}
\newcommand{  \act}{f}
\renewcommand{\mag}{{\bf p}}
\newcommand{  \ds }{d_s}
\newcommand{  \vd }{{\cal D}}
\newcommand{\brho}{{\bf f}}
\newcommand{  \g }{{\bf g}}
\newcommand{\f}{{\bf f}}
\newcommand{  \Dx }{D_\textnormal{E}}
\newcommand{  \Dt }{D_\textnormal{O}}
\newcommand{ \vo  }{v_0}
\newcommand{ \lat  }{{\TT^2_N}}
\newcommand{ \bconfig }{ \boldsymbol{ \hat \eta} }

\section{Introduction}
\label{sec:into}
Active matter systems are composed of interacting agents that consume energy from their environment. They exhibit a rich variety of collective phenomena including flocking~\cite{vicsekCollectiveMotion2012}, lane formation \cite{burgerLaneFormationSideStepping2016,bacikLaneNucleationComplex2023}, clustering \cite{buttinoniDynamicalClusteringPhase2013,filyAthermalPhaseSeparation2012,rednerStructureDynamicsPhaseSeparating2013} and pattern formation \cite{barSelfPropelledRodsInsights2020}. Active matter systems can be found in many natural and synthetic contexts \cite{vicsekCollectiveMotion2012}, ranging from biological systems such as bacteria \cite{sokolovPhysicalPropertiesCollective2012,bergRandomWalksBiology1993}, microtubules \cite{suminoLargescaleVortexLattice2012} and animal groups \cite{giardinaCollectiveBehaviorAnimal2008}, to physical and chemical systems such as colloids \cite{buttinoniDynamicalClusteringPhase2013} and self-propelled rods \cite{barSelfPropelledRodsInsights2020}. 
Several mathematical models have been developed using individual or agent-based approaches to capture the essential features of these systems. These models can help us understand the emergence of complex behaviours from simple rules at the microscopic level.

An important class of active matter models is based on self-propelled particles.  In some of these models, the propulsion directions of nearby particles tend to align with each other, which naturally leads to flocking behaviour~\cite{vicsekCollectiveMotion2012}.  Without such alignment, one often observes
a remarkable phenomenon of spontaneous condensation~\cite{catesMotilityInducedPhaseSeparation2015}, occurring in the absence of attractive forces between particles.  This behaviour, called Motility-Induced Phase Separation (MIPS), arises from the interplay between particle crowding and velocity reduction. When particles cluster together due to their self-propulsion, they also reduce their speed due to collisions and interactions. This creates a positive feedback loop that enhances the clustering and leads to a separation between a dense liquid-like phase and a dilute gas-like phase. MIPS has been extensively reviewed in \cite{catesMotilityInducedPhaseSeparation2015}. 
It is a striking example of how active matter can display emergent collective behaviours that are different from those of passive particles.

Numerical simulations have proven extremely valuable for understanding the emergence of MIPS and other complex phenomena in active matter models \cite{stenhammarActivityInducedPhaseSeparation2015,filyAthermalPhaseSeparation2012,solonGeneralizedThermodynamicsMotilityinduced2018,wittkowskiNonequilibriumDynamicsMixtures2017,brunaPhaseSeparationSystems2022,wysockiPropagatingInterfacesMixtures2016}. Analytical results can complement such studies by providing exact and general insights into this behaviour, but they are challenging to obtain, due to the non-equilibrium and nonlinear nature of active matter dynamics. 
One possible approach is to analyse coarse-grained models that capture the essential features of active matter systems at a macroscopic level, for example through density and velocity fields. Several phenomenological coarse-grained models have been proposed in the literature~\cite{bialkeMicroscopicTheoryPhase2013,catesMotilityInducedPhaseSeparation2015,speckDynamicalMeanfieldTheory2015,tailleurStatisticalMechanicsInteracting2008,wittkowskiScalarF4Field2014,brunaPhaseSeparationSystems2022, brunaDerivationMacroscopicModel2023}. Such models have yielded important insights, but deriving them from microscopic models relies on various assumptions and approximations, that are hard to justify.

A more rigorous and systematic approach is to derive coarse-grained descriptions directly from models of interacting particles~\cite{kourbane-housseneExactHydrodynamicDescription2018,erignouxHydrodynamicsActiveMatter2021}.
%One possible way to achieve this goal is to 
This can be achieved via the theory of hydrodynamic limits, in which macroscopic evolution equations are derived directly from microscopic stochastic rules.  It is most naturally applied to lattice gas models, where particles hop randomly on a discrete lattice according to some local rules. By taking a limit where the lattice spacing tends to zero, one can obtain continuum equations that exactly describe the large-scale behaviour of the system. This method has been successfully applied to various classes of lattice gas models \cite{kipnisScalingLimitsInteracting1998,quastelDiffusionColorSimple1992,erignouxHydrodynamicsActiveMatter2021} and can provide valuable insights into the physics of active matter \cite{kourbane-housseneExactHydrodynamicDescription2018}.

Following the recent interest in such lattice models of active systems \cite{kourbane-housseneExactHydrodynamicDescription2018,erignouxHydrodynamicLimitActive2021, brunaPhaseSeparationSystems2022},  this paper considers an {\model} ({\macc})  where particles interact via exclusion, with diffusive dynamics for their orientations.  We show that the macroscopic limit of {\macc} can be described exactly by an integro-differential equation. To explore the onset of MIPS, we analyse the stability of the homogeneous stationary state, in the parameter space of occupied volume fraction and P{\'e}clet number, through linear stability analysis and numerical simulations of both the microscopic and macroscopic models.

\subsection{Previous results}

A variety of active lattice gas models have been studied in recent years \cite{kourbane-housseneExactHydrodynamicDescription2018,erignouxHydrodynamicLimitActive2021,brunaPhaseSeparationSystems2022}.  In these models, each particle is equipped with an orientation variable, which determines the direction of its self-propulsion.  In the active lattice gas model of \cite{kourbane-housseneExactHydrodynamicDescription2018}, the orientation can point in a finite number of drift directions, aligned with the lattice axes.  In the terminology of~\cite{kipnisScalingLimitsInteracting1998}, that model is of \emph{gradient} type (see Definition \ref{def_grad_type}), and the macroscopic model can be derived exactly and explicitly. This enables a derivation of the phase diagram, including the limit of stability of the homogeneous state, and the density profiles for systems with inhomogeneous steady states. The macroscopic description of a more complex alignment model, with continuous particle orientation, was proven in~\cite{erignouxHydrodynamicLimitActive2021}: the latter model is of non-gradient type, which makes the analysis more challenging~\cite{guoNonlinearDiffusionLimit1988,quastelDiffusionColorSimple1992}.

In this work, we consider a different two-dimensional model where particles' orientations are unit vectors that undergo diffusive motion on a torus, similar to off-lattice models of active Brownian particles~\cite{filyAthermalPhaseSeparation2012,rednerStructureDynamicsPhaseSeparating2013}.  
This model was proposed in~\cite{brunaPhaseSeparationSystems2022}, where an approximate macroscopic description was derived within a mean-field approximation.  The resulting behaviour was compared with other coarse-grained off-lattice models, including a characterisation of the linear stability around a homogeneous state and the onset of MIPS.
They found that short-range interactions were important for observing MIPS because a similar model with long-range interactions did not exhibit phase separation.  In the following, we will characterise the exact hydrodynamic limit for this model without relying on the mean-field assumption.   Using this description, we revisit the stability of the homogeneous state and the onset of MIPS.  We verify the accuracy of this macroscopic description via a comparison with numerical simulations of the underlying microscopic models.

\subsection{Structure of the article}

The structure of the paper is as follows. 
In Section \ref{sec_notation_results}, we define the microscopic {model}, present its hydrodynamic limit, and summarise our main results for the stability of the homogeneous state.   
The linear stability analysis is detailed in Section \ref{sec_lin_stab}.
In Section \ref{sec_numerics}, we present numerical computations of both the macroscopic and microscopic models.  We compare the linear stability computation with the numerical simulations and demonstrate excellent agreement between simulations of the microscopic model and numerical solutions of the macroscopic equation.
The mathematical proof of the hydrodynamic limit is very technical and follows closely the derivation of \cite{erignouxHydrodynamicLimitActive2021}: we sketch the main arguments of the proof in Section \ref{sec_hydrolim}, focusing on the steps where our derivation differs from that of~\cite{erignouxHydrodynamicLimitActive2021}.

\section{Model definition and summary of main results}\label{sec_notation_results}

\subsection{Microscopic model: an {\model}}\label{sec_micro_model}

The model that we consider is an {\model} ({\macc}).  It is defined on a two-dimensional $N \times N$ periodic lattice
\begin{equation}
	\lat = \{ 1, \cdots, N\}^2.
\end{equation}
We define the \emph{occupation} configuration $\eta \in \{ 0,1 \}^\lat$. Each site ${\z}$ of $\lat$ is either occupied by a particle with orientation $\theta \in \SS = [0, 2 \pi)$, with periodic boundary conditions, ($\eta_{\z} =1$, $\theta_{\z} = \theta$) or empty ($\eta_{\z} =0$, $\theta_{\z} = 0$ by convention). The \emph{site} configuration at ${\z}$ is  ${\hat \eta}_{\z}:=(\eta_{\z}, \theta_{\z})$. The set of all configurations is given by 
\begin{equation}\label{equ_configs}
	\Sigma_N = \Big\{   (\eta_{\z}, \theta_{\z})_{{\z} \in \lat} \in ( \{0,1 \} \times \SS)^\lat \; \Big\vert \;\theta_{\z} = 0 \textnormal{ if } \eta_{\z} =0 \Big\}.
\end{equation}
Each point ${\z} \in \lat$ corresponds to a macroscopic point $\frac{{\z}}{N} \in \TT^2$ in the unit torus, on which the macroscopic profile will later be defined.

\medskip

The initial configuration %${{{\hat \eta}}} (0)$ 
is chosen in a state of local equilibrium, close to a smooth profile ${\hat \zeta}: \TT^2 \times \SS \to \RR_+$. More precisely, each site ${\z} \in \lat$ is initially occupied by a particle with probability $\zeta({\z}/N):=\int_\SS {\hat \zeta}(z/N,\theta) \dd \theta$. (For this construction to make sense, we assume that $\hat \zeta$ is such that $\zeta(x) \in [0,1]$ for all $x\in\TT^2$). The angle $\theta_{\z}$ is then sampled according to the probability distribution ${\hat \zeta}({\z}/N,\theta) \dd\theta/\zeta({\z}/N)$. We denote by $\mu^\star_{N,{\hat \zeta}}$ this initial distribution for our \macc.

\medskip

Informally, the dynamics of particles in the {\macc} can be described as follows
\begin{enumerate}[$\quad$--]
	\item Each particle attempts to perform a nearest neighbour random walk, weakly biased in the direction of their orientation. In particular, a jump in the $\u$ direction is attempted at rate $N^2 \Dx  + N \frac{\vo}{2}  ( \u \cdot {{{ \boldsymbol{e}_{\theta} }}})$, where  ${{{ \boldsymbol{e}_{\theta} }}} = (\cos(\theta), \sin(\theta) )^T$ is the drift direction, $\Dx$ is the spatial diffusion constant and $\vo$ is the self-propulsion speed.
	\item If the target site of a jump is occupied, the jump is aborted; if the site is otherwise empty, the jump is executed. This is called the \emph{exclusion rule}.
	\item Each particle's orientation diffuses according to a Brownian motion on $\SS$ with diffusion constant $\Dt$.
\end{enumerate}
The $N$-dependence of the particles' hop rates corresponds to a parabolic scaling \cite{kipnisScalingLimitsInteracting1998} and enables the exact characterisation of the hydrodynamic limit \cite{erignouxHydrodynamicLimitActive2021}.  It also ensures that for $N\to\infty$, a single particle's motion converges to an Active Brownian particle with diffusion constant $\Dx$, speed $\vo$ and angular diffusion $\Dt$.  The physical interpretation of this scaling for the hydrodynamic limit of interacting particles is discussed in Sec.~\ref{subsec_prop_hydro}, below.

To formalise the model, we identify the generator of this stochastic process, which can be decomposed
%The complete dynamics of the {\macc} can therefore be split 
into three parts, according to the $N$-dependence of the transition rates.  This generator is
%The symmetric and asymmetric contributions of the exclusion process, and the diffusion of particle orientations, each of those has a different scaling in the scaling parameter $N$. 
%The {\macc} is thus driven by the Markov generator

\begin{equation}\label{equ_markov_gen}
	L_N = N^2 {\cal L}_\textnormal{S} + N {\cal L}_\textnormal{A} + {\cal L}_\textnormal{O} ,
\end{equation}
whose elements we now define. The generator is defined on
\begin{equation}
	\mathcal{D}(L_N) = \Big\{g \in L^2( \Sigma_N) \;\vert \;\p^2_{\theta_{\z}} g \in L^2( \Sigma_N) \Big\}.
\end{equation} 
The nearest-neighbour simple symmetric exclusion process generator ${\cal L}_\textnormal{S}$ is defined as
\begin{equation} \label{equ_sym_gen}
	{\cal L}_\textnormal{S} g ({{{\hat \eta}}}) = \Dx \sum_{ \substack{ {\z},{\z}' \in \lat \\ \vert {\z}-{\z}' \vert =1} } 
	 \eta_{\z} (1- \eta_{{\z}'} ) [ g( {{{\hat \eta}}}^{{\z},{\z}'} ) - g ({{{\hat \eta}}}) ].
\end{equation}
Similarly, the asymmetric exclusion operator ${\cal L}_\textnormal{A}$ is:
\begin{equation}
	{\cal L}_\textnormal{A} g ({{{\hat \eta}}}) = \frac{\vo}{2} \sum_{ \substack{ {\z},{\z}' \in \lat \\ \vert {\z}-{\z}' \vert =1} }
	 ({\z}' -{\z}) \cdot \e_{\theta_{\z}}  \eta_{\z} (1- \eta_{{\z}'} ) [ g( {{{\hat \eta}}}^{{\z},{\z}'} ) - g ({{{\hat \eta}}}) ],
\end{equation}
%where ${{{ \boldsymbol{e}_{\theta} }}} = (\cos(\theta), \sin(\theta) )^T$. 
On its own, ${\cal L}_\textnormal{A}$ is not a Markov generator because it has negative jump rates. However, when added to the symmetric part, $N^2 {\cal L}_\textnormal{S} + N {\cal L}_\textnormal{A}$ becomes well-defined for $N$ large enough. All numerical simulations in this work satisfy this condition. In the expressions above, ${\hat \eta}^{{\z},{\z}'}$ denotes the configuration where the occupation variables ${{{\hat \eta}}}_{\z}$ and ${{{\hat \eta}}}_{{\z}'}$ have been exchanged in ${{{\hat \eta}}}$ 
\begin{equation}
	{{{\hat \eta}}}^{{\z},{\z}'}_y = \begin{cases}
		{{{\hat \eta}}}_{{\z}'} & \textnormal{ if } y = {\z},\\
		{{{\hat \eta}}}_{\z} & \textnormal{ if } y = {\z}',\\
		{{{\hat \eta}}}_y & \textnormal{ otherwise. } 
	\end{cases}
\end{equation}
Finally, ${\cal L}_\textnormal{O}$ is the generator for the diffusion of the orientation of each particle
\begin{equation}
\label{eq:orientationdiffusion}
	{\cal L}_\textnormal{O} g ({{{\hat \eta}}}) = \Dt \sum_{{\z}  \in \lat} \eta_{\z} \p^2_{\theta_{\z}} g ({{{\hat \eta}}}).
\end{equation}
Together, the initial distribution $\mu^\star_{N,{\hat \zeta}}$ and the Markov generator $L_N$ uniquely define a Markov process $({\hat \eta}(t))_{t\geq 0}$, whose macroscopic behaviour we want to characterise.

\subsection{Hydrodynamic Limit of the {\macc}}

We now introduce the exact hydrodynamic limit of the orientation density for the {\macc}, whose derivation is outlined in Section \ref{sec_hydrolim}. 
%This hydrodynamic equation is exploited in Sections~\ref{sec_lin_stab} and \ref{sec_numerics} to analyse MIPS in the \macc\ via a linear stability analysis of its homogeneous solution, which is compared with a numerical analysis of stability.

The macroscopic equation of the \macc\ depends on the \emph{self-diffusion coefficient} of a tagged particle in a simple symmetric exclusion process (SSEP).  (The same dependence appears in the models studied in~\cite{erignouxHydrodynamicLimitActive2021}.)

\begin{definition}[Self-diffusion coefficient]
In $\ZZ^d$, consider an infinite SSEP and a tracer particle placed initially at the origin. We place a particle at each site in $\ZZ^d \setminus \{ 0 \}$ independently with probability $\rho$. The position of the tracer at time $t$ is $X_t \in \mathbb{Z}^d$ and we denote by $Q_\rho$ the distribution of the resulting process. Then the self-diffusion coefficient is 
	\begin{equation}
		d_s( \rho ) = \lim_{t\to 0} \EE_{Q_\rho} \Bigg( \frac{ \vert X_t \vert^2 } { 2t} \Bigg).
	\end{equation}
\end{definition}
Spohn \cite{spohnTracerDiffusionLattice1990} obtained a variational formula for the self-diffusion coefficient. The function $d_s$ was shown to be Lipschitz continuous by Varadhan in \cite{varadhanRegularitySelfDiffusionCoefficient1994} for $d \geq 3$ and was later shown to be of class $C^\infty$ for $d \geq2$ by Landim, Olla and Varadhan in \cite{landimSymmetricSimpleExclusion2001}.
Various approximations to $d_s$ are discussed in Section~\ref{subsec_prop_hydro}, below.

The hydrodynamic equation describes the evolution of the local macroscopic density of particles with angle $\theta$, which we denote by $f(t,\x,\theta)$.  It can be interpreted formally as
\begin{equation}
\label{eq:ftx}
f(t,\x,\theta) \dd\theta \approx \frac{1}{(2 N \ep +1)^2}\sum_{\vert {\z}-N\x\vert \leq N \ep} \eta_{\z}(t) {\bf{1}}_{\{\theta_{\z}(t)\in [\theta, \theta+ \dd\theta] \}}.
\end{equation}
where $\varepsilon$ is the side length of a small macroscopic box, in which we measure particle density, where $1 \gg \ep \gg 1/N  \gg d\theta$.

We will sometimes change our notation to $f_t(\x,\theta)$ or 
$f_\theta(t,\x)$ to emphasise the dependency on one of the variables (see e.g. Definition \ref{def_weak_sol} and remark \ref{rem:Einstein} below). We also define the polarisation $\mag $ and the particle density $\rho$, by
\begin{equation}
	\mag(t,\x) = \int_0^{2\pi} {{{ \boldsymbol{e}_{\theta} }}} \act(t, \x, \theta) \dd \theta , \qquad \qquad \rho(t,\x):= \int_0^{2\pi} \act(t, \x, \theta) \dd \theta .
 \end{equation}
 [Recall that ${{{ \boldsymbol{e}_{\theta} }}} = (\cos(\theta), \sin(\theta) )^T$ is the drift direction of a particle with orientation $\theta$.]  It is also convenient to introduce two coefficients $\vd$ and $s$ as
 \begin{equation}
 \vd (\rho ) = \frac{1- \ds (\rho)}{\rho} ,\qquad \mbox{ and } \qquad s(\rho) = \vd(\rho ) - 1.
 %\label{equ_coefficients}
\end{equation}
With these definitions, the macroscopic evolution equation for the density field $f$ of the \macc\ defined in \eqref{eq:ftx} can be derived rigorously from the microscopic model as
\begin{equation}
\label{act_pas_eq}
		\pt \act = \Dx \div \Big[ \ds( \rho ) \nabla \act+ \act \vd (\rho ) \nabla \rho \Big] - \vo \div \Big[ \act s(\rho ) \mag   + {{{ \boldsymbol{e}_{\theta} }}} \ds ( \rho ) \act \Big]  + \Dt \p_\theta^2 \act. 
\end{equation}

%As mentioned previously, an analogous integro-differential equation has been derived from a microscopic active lattice gas in \cite{erignouxHydrodynamicLimitActive2021}. 
In the simpler models studied in \cite{kourbane-housseneExactHydrodynamicDescription2018}, particle jumps are only biased in two directions. However, a straightforward generalisation to a continuum of directions would also lead to integro-differential equations similar to \eqref{act_pas_eq}, but in a simpler form due to the gradient nature of the models considered there. Similar integro-differential equations have been studied in the context of flocking models (see e.g. \cite{barbaroPhaseTransitionsKinetic2016} and references therein), but unlike in our case, where the evolution equation can be derived rigorously from the microscopic model, deriving such evolution equations typically relies on mean-field assumptions, meaning particle's interactions are averaged over a neighbourhood.
% containing many particles.}

\medskip

Note that to make this hydrodynamic limit as easy to read as possible, we stated it here in the informal form \eqref{act_pas_eq} without regard for the well-posedness of the equation and without proper justification for the convergence result \eqref{eq:ftx}. The rigorous statement of the \macc's hydrodynamic limit involves the process's empirical measure $ \pi_t^N$, defined by its integral against test functions $H$
\begin{equation}
\label{eq:empmeasH}
\langle \pi^N_t , H \rangle:=\frac{1}{N^2}\sum_{{\z}\in \lat}\eta_{\z}(t) H\left(\frac{{\z}}{N},\theta_{\z}(t)\right).
\end{equation}
We postpone the full statement of the hydrodynamic limit to Section \ref{sec_hydrolim}, where we give a precise definition of what we mean by solutions of \eqref{act_pas_eq}, and state in Theorem~\ref{thm_main} that the empirical measure converges in probability to the local density field \eqref{eq:ftx}, namely
\begin{equation}
\pi_t^N(\dd \x ,\dd\theta)\underset{N\to\infty}{\longrightarrow} f(t, \x ,\theta) \dd \x \dd \theta.
\end{equation}

The proof of the hydrodynamic limit for the \macc\ (Theorem \ref{thm_main}) involves severe mathematical challenges~\cite{erignouxHydrodynamicLimitActive2021}, which mostly arise because the model is of non-gradient type (See Definition \ref{def_grad_type}). % in the terminology of~\cite{kipnisScalingLimitsInteracting1998}.  
However, the \macc\ considered here is very close to those analysed in~\cite{erignouxHydrodynamicLimitActive2021}: the only significant difference is that the orientation dynamics is a diffusion on $\SS$ instead of a Markov jump process.  While this means that the hydrodynamic limit proofs of \cite{erignouxHydrodynamicLimitActive2021} cannot be used \emph{verbatim} in this case, the proof of Theorem~\ref{thm_main} is a straightforward generalisation of that work.  The main steps of the proof are described in Sec.~\ref{sec_hydrolim}, and the aspects that differ from \cite{erignouxHydrodynamicLimitActive2021} are explained.  (The diffusive orientational dynamics simplifies some aspects of the proof, see Remark \ref{rem:angulardiffusion})
Whilst the hydrodynamic limit is an extension of \cite{erignouxHydrodynamicLimitActive2021}, numerical and analytical analysis of \eqref{act_pas_eq} is hindered by the density dependence of the diffusion constant: this enters via $d_s(\rho)$, for which no explicit formula is available. As noted in \cite{erignouxHydrodynamicLimitActive2021} the proof of the hydrodynamic limit can easily be extended to higher dimensions $d \geq 2$. 

To understand the physical structure behind the hydrodynamic equation \eqref{act_pas_eq}, we observe that if the particles are passive, $v_0=0$, the microscopic model is a reversible Markov chain and its invariant measure is a product over the lattice sites.  In this case, we explain in Remark~\ref{rem:GradFlow} that its free energy is
\begin{equation}
S[\act]=\int_{\TT^2} \left\{ [1-\rho(\x)]\log[1-\rho(\x)]+\int_{\SS} \act(\x, \theta)\log [2\pi \act(\x, \theta)] \dd\theta \right\} \dd \x
\label{equ:free-energy}
\end{equation}
We also define a parametric distribution which plays the role of a mobility
\begin{equation}
\label{eq:sigma1}
\sigma_{f,\theta}(\dd\theta')= s(\rho ) f(\theta) f({\theta'}) \dd \theta' + \ds(\rho) f(\theta) \delta_{\theta}(\dd \theta').
\end{equation}
where $f(\theta):=f(\cdot,\cdot,\theta)$ and $\delta_\theta(\dd \theta')$ is the Dirac distribution at $\theta$ on $\SS$, see Remark \ref{rem:Einstein} for a further discussion.
Then the hydrodynamic equation \eqref{act_pas_eq} can be factorised as
\begin{equation}
\label{eq:GFv0}
    \pt f(t, \x, \theta) = \nabla \cdot \int \left[ \Dx \nabla \frac{\delta S}{\delta f}(t, \x, \theta') + v_0 \boldsymbol{e}_{\theta'} \right] \sigma_{f,\theta}(d\theta') + \p_\theta \left[ \Dt f \p_\theta \frac{\delta S}{\delta f}(t, \x, \theta) \right]
\end{equation}
In the passive case $v_0=0$, this is a gradient-flow for $S$, as one would expect for the hydrodynamic limit of a microscopic model with time reversal symmetry \cite{mielkeRelationGradientFlows2014}.  In general, the objects in square brackets are thermodynamic forces that drive the evolution of the model, note that the active self-propulsion is not the gradient of any free energy, consistent with the fact that the microscopic model has a non-equilibrium steady state.

This form for the hydrodynamic equation is discussed further in Remark \ref{rem:GradFlow}.

\subsection{Properties of the hydrodynamic equation}
\label{subsec_prop_hydro}

The macroscopic evolution described by \eqref{act_pas_eq} depends on five parameters. Three of them appear in~\eqref{act_pas_eq} itself: the spatial diffusion constant $\Dx$, self-propulsion speed $\vo$, and angular diffusion constant $\Dt$. There are also two further parameters: the domain size, and the occupied volume fraction of the active particles. 

The lattice spacing (or `particle size') in the microscopic model has already been taken to zero in a domain of fixed size.
The parameters of the model encode two additional length scales 
\begin{align}
	\ell_p = \frac{\vo }{\Dt}, && \ell_D = \sqrt{\frac{\Dx}{\Dt}},
\end{align}
where $\ell_p$ (resp. $\ell_D$) represents the typical distance travelled by a particle because of its drift (resp. because of its symmetric motion) before its angle reorients significantly.
For the model of \cite{kourbane-housseneExactHydrodynamicDescription2018}, the widths of the interfaces separating the dense and dilute phases of MIPS are proportional to $\ell_D$.

For a physical interpretation of the hydrodynamic limit, we emphasize once again that the $N$-dependence of the generator $L_N$ ensures that $\ell_p$ and $\ell_D$ are of order unity as $N\to\infty$, while the range over which particles interact by exclusion is tending to zero.  In physics, it is more common to take $\ell_p$ as a fixed multiple of the particle size~\cite{tailleurStatisticalMechanicsInteracting2008,filyAthermalPhaseSeparation2012,rednerStructureDynamicsPhaseSeparating2013}, on the grounds that both length scales are intrinsic properties of individual particles. In this case, various approximate descriptions are available for the (fluctuating) hydrodynamic behaviour~\cite{tailleurStatisticalMechanicsInteracting2008,wittkowskiNonequilibriumDynamicsMixtures2017,wysockiPropagatingInterfacesMixtures2016}, but the hydrodynamic limit cannot be characterised exactly.  Hence, our focus is on the generator $L_N$, for which exact computations are possible.

Overall, a natural set of dimensionless parameters that is sufficient to capture the behaviour of \eqref{act_pas_eq} consists of
\begin{align}
	\phi = \frac{1}{\vert \TT^2 \vert}\int_{\TT^2} \rho(t,\x) \dd \x , && \textnormal{Pe} = \frac{\ell_p}{\ell_D}, && \ell = \frac{\ell_D}{\sqrt{\vert \TT^2 \vert}},
\end{align}
which are the occupied volume fraction, the P{\'e}clet number and the diffusive length scale respectively.
In the following, the domain size is set to unity, %$|\TT^2|=1$, this does not lose any generality since it can always be achieved by a rescaling of the spatial coordinate. Similarly, we take $\Dx=1$ since this can always be achieved by 
rescaling time as $t' = t/\Dx$. In dimensionless parameters, \eqref{act_pas_eq} becomes
\begin{equation}
\label{act_pas_eq_2}
		\ell^2 \pt \act = \ell^2 \div \Big[ \ds( \rho ) \nabla \act+ \act \vd (\rho ) \nabla \rho \Big] - \ell \text{Pe} \div \Big[ \act s(\rho ) \mag   + {{{ \boldsymbol{e}_{\theta} }}} \ds ( \rho ) \act \Big]  + \p_\theta^2 \act. 
\end{equation}

The analysis of \eqref{act_pas_eq} is hindered by the density dependence of the diffusion constant: this enters via $d_s(\rho)$, for which no explicit formula is available.  However, previous work has led to very accurate numerical approximations.  Our approach here is to approximate $d_s$ by a polynomial derived in~\cite{masonMacroscopicBehaviourTwoSpecies2023}, which is asymptotically exact at both low and high density. This approximation -- defined in \eqref{eq_ds_approx} below -- is used for all numerical computations in the following. An alternative asymptotically accurate formula was derived in \cite{nakazatoSiteBlockingEffect1980}; other approximations are available as mean-field-type approximations \cite{illienNonequilibriumFluctuationsEnhanced2018,rizkallahMicroscopicTheoryDiffusion2022}, variational bounds~\cite{aritaGeneralizedExclusionProcesses2014,aritaBulkDiffusionKinetically2018} or low-rank tensor approximations~\cite{dabaghiComputationSelfdiffusionCoefficient2022,dabaghiTensorApproximationSelfdiffusion2023}. Our polynomial approximation is employed for its accuracy at high and low density and ease of use. Other approximations, such as mean-field \cite{brunaPhaseSeparationSystems2022}, lead to similar qualitative behaviour, but the quantitative predictions for the onset of MIPS are less accurate. 

\begin{figure}
\centering
	\includegraphics[width=.6\textwidth]{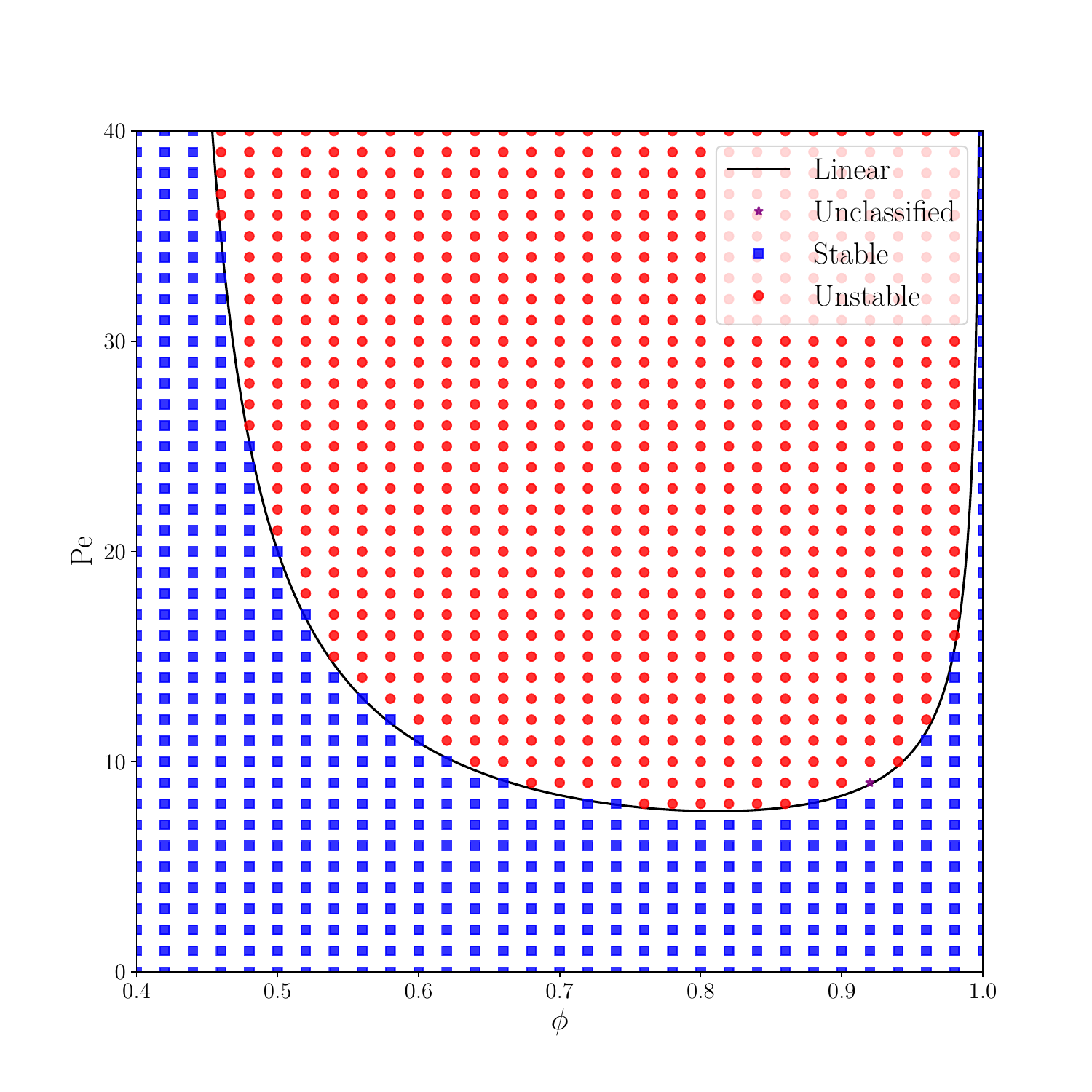}
	\caption{%
Illustration of the linear stability boundary~\eqref{equ_pert} for the homogeneous solution of \eqref{act_pas_eq} (solid line), compared with numerical solutions of the equation.  Results are shown for the illustrative parameter value $\ell=0.5$
 The coloured points indicate whether the numerical results indicate stable or unstable behaviour. The comparison between analytical and numerical results is discussed in the main text.
	}
	\label{fig_1}
\end{figure}

With this numerical approximation of $\ds$ in hand, we have analysed MIPS in our \macc\ in two different ways. In Section~\ref{sec_lin_stab}, we study the linear stability of the homogeneous solution to \eqref{act_pas_eq}.  A linear instability at this level signals that MIPS is taking place.  A detailed discussion is given below, but we summarise the behaviour in Fig.~\ref{fig_1}, which shows the onset of linear stability as a function of Pe and volume fraction $\phi$. This is compared with the behaviour of numerical solutions of~\eqref{act_pas_eq}.  For low densities, we observe that the system is always linearly stable.
For higher densities, the onset of instability in the numerical solutions coincides with the boundary of linear stability. %For moderate densities ($\phi\lesssim0.8$), the onset of instability in the numerical solutions coincides with the limit of linear stability.  However, for higher densities, the linear stability analysis does not predict the instabilities observed numerically.  We attribute this effect to nonlinear instabilities that set in when the lattice becomes almost fully occupied.

In addition to these results for stability, Section~\ref{sec_numerics} presents numerical solutions of the macroscopic equation \eqref{act_pas_eq}, which is compared with particle-based stochastic simulations of the microscopic model.  The agreement is excellent, including in the MIPS regime: this is consistent with the rigorous derivation of the hydrodynamic limit, combined with the numerically accurate approximation for $d_s$.

\section{Linear Stability}\label{sec_lin_stab}

\begin{figure}
\centering
	\includegraphics[width=.6\textwidth]{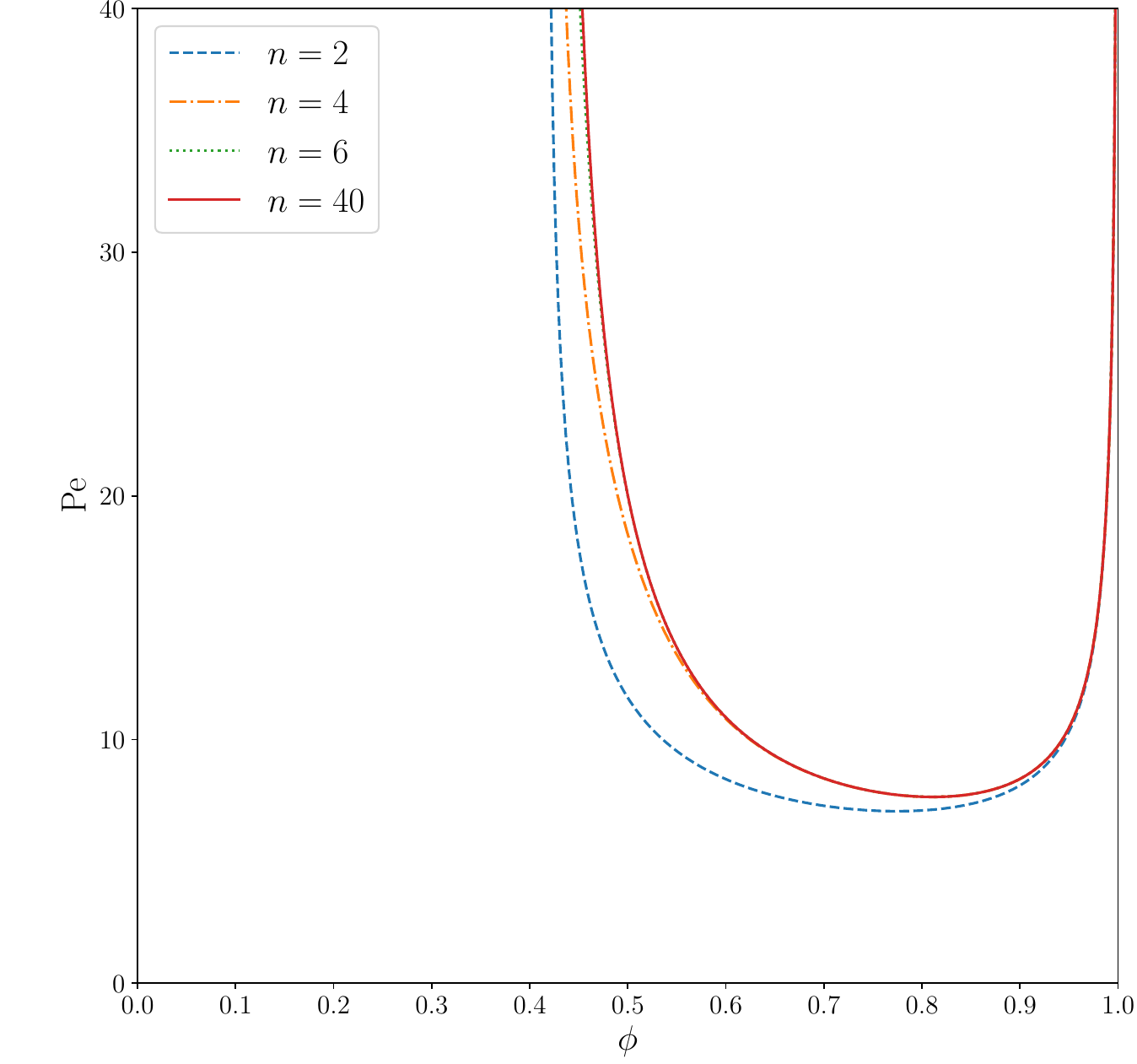}
	\caption{Plots of $\Re{\lambda_n^\textnormal{max}} =0$ as determined from \eqref{equ_lambda_n}. As $n \to \infty$, these curves converge to the boundary between stability and instability of the linearised problem \eqref{linear_model}. 
The case $n=6$ is indistinguishable from $n=40$ due to the rapid convergence of the scheme.
	The area above the limiting curve corresponds to the linear problem having a positive eigenvalue. We take the representative parameter value $\ell = 0.5$. }
	\label{fig_2}
\end{figure}

We consider the linearisation of equation \eqref{act_pas_eq} around the homogeneous steady state $\act_* \equiv \frac{\phi}{2 \pi}$. We insert $\act= \act_* + \delta \tw{\act}$, and $\rho = \phi + \delta \tw{\rho}$, for $\delta \ll 1$. Furthermore $\int {{{ \boldsymbol{e}_{\theta} }}} \dd \theta =0$ and therefore $\mag=\mag[f] = \delta \tw{\mag}:= \mag[ \delta\widetilde{f}]$. The resulting linearised problem is:
\begin{align}\label{linear_model}
		\pt \tw{\act} &= \Dx \div \Big[ \ds( \phi ) \nabla \tw{\act} + \frac{\phi}{2 \pi} \vd(\phi) \nabla \tw{\rho}  \Big] 
		\\ \nonumber
		& \quad \quad \quad \quad \quad - \vo \div \Big[ 
		\frac{\phi}{2 \pi} s(\phi ) \tw{\mag}   
		+ \ds^\prime ( \phi )   {{{ \boldsymbol{e}_{\theta} }}} \frac{\phi}{2 \pi} \tw{\rho} 
		+ {{{ \boldsymbol{e}_{\theta} }}} \ds ( \phi ) \tw{\act}  \Big]  
		+ \Dt \p_\theta^2 \tw{\act}.
\end{align}
The linear stability analysis considers solutions of the form $\tilde f \propto {\rm e}^{\lambda t}$, in which case \eqref{linear_model} becomes an eigenproblem.
Since the domain is periodic,
the eigenfunctions take the general form
\begin{align}\label{equ_pert_general}
	\tw{f} (t,\x,\theta)  = \sum_{ \boldsymbol{\omega} \in 2 \pi\NN^2} e^{\lambda t + i \boldsymbol{\omega} \cdot \x} \sum_{k\geq 0 } \Big[A_{k,\boldsymbol{\omega}} \cos( k \theta)+B_{k,\boldsymbol{\omega}} \sin( k \theta)\Big].
\end{align}
In the spatial variable, the equation is closed at each frequency, and therefore we can consider each spatial frequency $\boldsymbol{\omega}$ separately. To find the eigenvalue with the largest real part, we seek the lowest non-trivial frequency in $x$, so it is sufficient to consider $\boldsymbol{\omega} = (\omega,0) = (2\pi, 0)$, which corresponds to a wave along one axis.  Furthermore, taking a wave in the $x_1$ direction, equation \eqref{linear_model} immediately yields $B_k = 0$. Hence it is sufficient to consider
\begin{align}\label{equ_pert}
	\tw{f} (t,\x,\theta)  = e^{\lambda t + i \omega x_1} \sum_k A_k \cos( k \theta) .
\end{align}
Then \eqref{linear_model} becomes
\begin{subequations}\label{equ_difference}
\begin{align}
\lambda A_0
&= W_{11}A_0 + W_{12}A_1,
		\\
\lambda A_1
&= W_{21}A_0 + W_{22}A_1 + b A_2,
\\ \label{equ_diff}
\lambda A_k &= (a- \Dt k^2)A_k + b(A_{k-1}+A_{k+1}), \mathrm{ ~for~ } k\geq 2,
\end{align}
\end{subequations}
where $a = -\Dx \ds( \phi ) \omega^2$, $b = - \frac{\vo}{2} i \omega \ds( \phi ) $ 
and
\begin{subequations}\label{equ_diffference_matrix}
\begin{align}
	W_{11} &= a -\Dx \omega^2 \phi \vd(\phi) = -\Dx \omega^2
	,~&
	W_{12} &= b - \frac{\vo}{2} i \omega  \phi s(\phi ) = - \frac{\vo}{2} i \omega (1-\phi),
	\\
	W_{21} &= 2b - \vo i \omega  \phi \ds^\prime ( \phi )    	
	,~&
	W_{22} &= a  - \Dt.
\end{align}
\end{subequations}
For $k \geq 2$, \eqref{equ_diff} are the difference equations derived from the Mathieu Equation \cite{zienerMathieuFunctionsPurely2012}. In that case, \eqref{equ_diff} applies for all $k$, including $k=0,1$. In our case, there are additional terms at $k=0$ and $k=1$ due to the terms $\rho$ and $\mag$ in~\eqref{linear_model}, which are integrals of $f$. The simplicity of the underlying integrals means that these terms only appear for $k=0,1$, which means that we are still able to close the equations, as follows. Equation \eqref{equ_diff} has two linearly independent solutions \cite{gautschiComputationalAspectsThreeTerm1967}, one diverges and the other behaves as
\begin{equation}\label{equ_ratio}
	\frac{A_{k+1}}{A_k} \sim \frac{b}{a -k^2 \Dt - \lambda},
\end{equation}
as $k \to \infty$.
Hence a convergent solution to the full problem \eqref{equ_difference}, $(A_k)_{k\geq0} \in L^2$, only exists when $\lambda$ takes specific values, called the characteristic numbers. For numerical computation, it is useful to rewrite the recurrence relation as an eigenvalue problem of a countable tridiagonal matrix
\begin{equation}\label{equ_matrix_system}
	\begin{pmatrix}
\multicolumn{2}{c}{\multirow{2}{*}{\Large W}} 
& 0  & 0 & \dots \\
 & & b & 0 & \dots \\
 0  & b &  a -4 \Dt  & b & \dots \\
 0  & 0& b &  a -9 \Dt & \dots \\
\vdots & \vdots & \vdots & \vdots &\ddots
\end{pmatrix} 
\begin{pmatrix}
A_0  \\
A_1 \\
A_2 \\
A_3 \\
\vdots
\end{pmatrix} = \lambda \begin{pmatrix}
A_0  \\
A_1 \\
A_2 \\
A_3 \\
\vdots
\end{pmatrix},
\end{equation}
whose $W$ is defined in \eqref{equ_diffference_matrix}.
For each eigenvalue of \eqref{equ_matrix_system}, 
%there exist a sequence of eigenvalues $\lambda_n$ to the truncated tridiagonal $n \times n$ matrix satisfying 
there is a corresponding eigenvalue $\lambda_n$ of the corresponding truncated $n \times n$ matrix: the resulting sequence $(\lambda_n)_{n\in\mathbb{N}}$ behaves for large $n$ as
(cf. \cite{ikebeEigenvalueProblemInfinite1996})
\begin{equation} \label{equ_lambda_n}
	\lambda-\lambda_n = \frac{b A_{n} A_{n+1}}{\Vert A \Vert_2} (1 + o(1)) \leq 0.
\end{equation}
Combining with \eqref{equ_ratio} yields the bound $\vert \lambda-\lambda_n \vert \lesssim \left[\frac{C^n}{n!(n+1)!}\right]^2$. Thus we can approximate the eigenvalue with the largest real part $\lambda^\textnormal{max}$ by $\lambda^{\textnormal{max}}_n$ for large $n$. Figure \ref{fig_2} plots $\Re{\lambda^{\textnormal{max}}_n} =0$ for $n= 2, 4, 6, 40$, displaying the rapid convergence to $\Re{\lambda^{\textnormal{max}}} =0$, the boundary between stability and instability of the linearised problem \eqref{linear_model}.  

\begin{remark}[Sharp-interface limit] In the limit $\ell \ll 1$, the inhomogeneous steady states of \eqref{act_pas_eq} consist of uniform regions separated by narrow interfaces. These are the phase-separated states that occur in MIPS, where it is conventionally assumed that the interfacial widths are much smaller than the sizes of the domains of the individual phases.
For a fixed $\omega$, $\Dx$ and $\phi$, let $\textnormal{Pe}^*(\ell)$ be the critical Peclet such that $\lambda^\textnormal{max} = 0$.
In the limit  $\ell \to 0$ assume that $\textnormal{Pe}^*(\ell) \to \textnormal{Pe}^*(0)$. Then $\ell^{-k} A_k(\ell) \to c_k$, and in particular \eqref{equ_difference} implies $ \det( W) \to 0$. Setting $ \lim_{\ell \to 0} \det(W) = 0$
 yields the limiting curve 
 \begin{equation}\label{equ_spinodal}
	{\textnormal{Pe}^*}{(0)}^2 = \frac{-1}{(1-\phi)(\ds(\phi) + \phi \ds^\prime(\phi))},
\end{equation}
which has asymptotes at $\phi = 1$ and when $\ds(\phi)+\phi \ds^\prime(\phi) = 0 $ for $\phi \in (0,1)$. 
The functions $\ds(\phi)$ and $\ds^\prime(\phi)$ are continuous and $\ds(0)>0$, hence $\ds(\phi)+\phi \ds^\prime(\phi)$ will be positive in some neighbourhood of $\phi = 0$. 
Therefore, for volume fractions below some critical level, the homogeneous steady state is stable for all $\textnormal{Pe}$. 
\end{remark}
\medskip
\begin{remark}[Run and tumble particles] 
    In the case of run and tumble particles, the diffusive dynamics of particle orientation is replaced by uniformly random reassignment at a given rate $\Dt$. In \eqref{act_pas_eq} $\Dt \p^2_\theta \act$ would be replaced by $\Dt (\frac{\rho}{2\pi} - \act) $. As explored in \cite{catesWhenAreActive2013}, run and tumble dynamics result in similar phenomenology to diffusive orientations. However, in the linear stability calculations above
     $ -k^2 \Dt \mapsto  -\Dt \id_{\{k>0\}}$ resulting in increased dependence on higher-order Fourier modes. 
\end{remark}
\medskip

Several comments should be borne in mind when interpreting the results of this linear stability analysis. For systems described by ordinary differential equations (with sufficient regularity), the boundary of linear and nonlinear stability are the same \cite[Sec. 9.3]{boyceElementaryDifferentialEquations2017}. They are also equivalent in quasilinear partial differential evolution equations where terms containing the highest order spatial derivatives are linear so that the regularising effect of the linear terms outweighs the nonlinear terms \cite[Sec. 4.3]{toddkapitulaSpectralDynamicalStability2013}, allowing the solution to be written explicitly using a semigroup generator. However, this does not apply to \eqref{act_pas_eq} because of the nonlinear terms involving second-order derivatives. We expect agreement between linear and nonlinear stability, but a rigorous proof is beyond the scope of this paper. Similarly, nonlinear terms can cause difficulty in rigorously proving exponential convergence to equilibrium for linearly stable cross-diffusion systems \cite{burgerNonlinearCrossDiffusionSize2010}.

\medskip

For the physical setting of MIPS, the limit of linear stability of the homogeneous state is known as the spinodal \eqref{equ_spinodal}.  In the unstable regime, Equation~\eqref{act_pas_eq} has a stable inhomogeneous steady-state solution, which corresponds to phase separation.  (In fact, there is a whole family of such solutions related by spatial translations.)  However, the physics of phase separation is more complicated than this, because one generally expects a parameter regime where \eqref{act_pas_eq} supports homogeneous and inhomogeneous steady solutions, which are both stable. The boundary of this region is known as the binodal \cite{catesMotilityInducedPhaseSeparation2015}. In this regime, the microscopic model exhibits metastability~\cite{bovierMetastabilityPotentialTheoreticApproach2015}: it supports fluctuations which can lead to spontaneous transitions between the metastable states.  The hydrodynamic description does not capture these; they would require an analysis of large deviations of the empirical measure \cite{kipnisScalingLimitsInteracting1998,quastelLargeDeviationsSymmetric1999,agranovEntropyProductionIts2022,erignouxHydrodynamicsActiveMatter2021}, which is beyond the scope of this work.

\section{Numerical Results}\label{sec_numerics}

\subsection{Numerical PDE scheme} \label{sec_pde_scheme}

We use a first-order finite-volume scheme to obtain numerical solutions to \eqref{act_pas_eq}.  We first rewrite the equation in the form
\newcommand{\divz}{{\nabla_\zeta \cdot}}
\begin{align} \label{equ_almost_gradient}
		\pt \act = - \div \Big[ (M^{x_1} U^{x_1}, M^{x_2} U^{x_2}) \Big] - \p_\theta M^\theta U^\theta \act,
\end{align}
where $M^{x_1}, M^{x_2}, M^\theta$ are scalar mobilities and ${\bf U} = ({U}^{x_1}, {U}^{x_2}, U^\theta)$ is the velocity vector. The mobilities are defined by $M^{x_1}=M^{x_2}=\act \ds$, $M^\theta=\act$ and the velocities are given by
\begin{align}
	(U^{x_1}, U^{x_2}) &= - \Dx \Big[ \nabla \log \act   + \nabla Q( \rho ) \Big] + \vo \Big[ {{{ \boldsymbol{e}_{\theta} }}} + \frac{ \mag s(\rho )}{ \ds(\rho) } \Big],
	\\
	U^\theta &= - \Dt \p_\theta \log \act,
\end{align}
where $Q: [0,1] \to \RR$ is such that $Q'(x) = \vd (x)/ \ds(x)$. We note that \eqref{equ_almost_gradient} is not a Wasserstein gradient flow as the velocity  ${\bf U}$ cannot be written as the derivative of an entropy (unless $\vo = 0$).

\medskip

The function $d_s$ is not known explicitly, so we approximate it by the polynomial approximation derived in \cite{masonMacroscopicBehaviourTwoSpecies2023} which is asymptotically exact to the first order at both low and high density:
\begin{equation}
\label{eq_ds_approx}
	\tw{\ds} (\rho) = (1-\rho)\Bigg( 1 - \alpha \rho + \frac{\alpha(2\alpha-1)}{2\alpha +1} \rho^2 \Bigg),
\end{equation}
where $\alpha = \pi/2 -1$.

We discretize the phase space $\Upsilon = [0,1]^2 \times [0, 2\pi)$ into $N_{x_1} \times N_{x_2} \times N_\theta$ boxes. Each cell $C_{i, j, k}$ has volume $\Delta {x_1} \Delta {x_2} \Delta \theta$ with centre $(i\Delta {x_1}, j \Delta {x_2}, k \Delta \theta)$. Furthermore, the time interval $[0,T]$ is discretized by $t = \Delta t, 2\Delta t,  \dots, T$. 
Define the cell averages
\begin{equation}
	f_{i,j,k}(t) = \frac{1}{ \Delta x_1 \Delta x_2 \Delta \theta } \int_{C_{i, j, k}} f(t, \x, \theta) \dd \x \dd \theta. 
\end{equation}
We use the finite-scheme 
\begin{align} \label{equ_ODE_system}
\frac{\dd}{\dd t} f_{i,j,k} = 
		&- \frac{F^{x_1}_{i+1/2,j,k}-F^{x_1}_{i-1/2,j,k} }{\Delta x_1} \\ \nonumber
		&- \frac{F^{x_2}_{i,j+1/2,k}-F^{x_2}_{i,j-1/2,k} }{\Delta x_2}
		- \frac{F^{\theta}_{i,j,k+1/2}-F^{\theta}_{i,j,k-1/2} }{\Delta \theta}.
\end{align}
for $ 1 \leq i \leq N_{x_1}, 1 \leq j \leq N_{x_2}, 1 \leq k \leq N_\theta$, where $F$ is the upwind flux
\begin{equation}
	F^{x_1}_{i+1/2,j} = (U^{x_1}_{i+1/2,j})^+ f_{i,j} + (U^{x_1}_{i+1/2,j})^- f_{i+1,j},
\end{equation} 
and similarly for $F^{x_2}$ and $F^{\theta}$. We discretize the integrals 
\begin{align}
	\rho_{i,j} = \Delta \theta \sum_k f_{i,j,k}, &  & \mag_{i,j,k} =  \Delta \theta \sum_k {\bf e}_{k \Delta \theta} f_{i,j,k},
\end{align}
where ${{{ \boldsymbol{e}_{\theta} }}} = (\cos(\theta), \sin(\theta) )^T$.
The velocities are approximated by centred differences or centred averages,
\begin{align}
			U^{x_1}_{i+1/2,j,k} =&
	- \Dx \Bigg[ 
	 \frac{\log f_{i+1,j,k} - \log f_{i,j,k} }{\Delta {x_1}}
	+\frac{Q( \rho_{i+1,j} ) - Q( \rho_{i,j} ) }{\Delta {x_1}}
	\Bigg]
	\\ \nonumber
	&+ \vo \Bigg[
	\frac{1}{2} \Bigg( \frac{ \mag_{i+1,j}  s( \rho_{i+1,j})  }{  \ds (\rho_{i+1,j}) } + \frac{ \mag_{i,j} s( \rho_{i,j})  }{  \ds (\rho_{i,j}) } \Bigg) + \e_{k \Delta\theta}   
	\Bigg],
\end{align}
and similarly for $U^{x_2}, U^{\theta}$. Finally, we discretize the system of ODE \eqref{equ_ODE_system} by the forward Euler method with an adaptive time stepping condition satisfying
\begin{equation}\label{equ_cfl}
	\Delta t \leq \min \Bigg\{ \frac{\Delta {x_1}}{a_{x_1}}, \frac{\Delta {x_2}}{a_{x_2}}, \frac{\Delta \theta}{a_\theta} \Bigg\},
\end{equation}
where $a_{x_1} = \max \{ \vert U^{x_1}_{i,j,k} \vert \}$, $a_{x_2} = \max \{ \vert U^{x_2}_{i,j,k} \vert \}$, and $a_\theta = \max \{ \vert U^{\theta}_{i,j,k} \vert \}$. In \cite{krukFiniteVolumeMethod2021}, they derive a CFL condition for a nonlocal model of active particles and show that it leads to a positivity-preserving numerical scheme. A key difference is that their scheme is second-order in phase space as they use a linear density reconstruction at the interfaces that preserves positivity. In contrast, we follow \cite{brunaPhaseSeparationSystems2022} and use the values at the centre of the cells instead. In our numerical tests (in which $\Delta t \leq 10^{-5}$), we observe \eqref{equ_cfl} to be sufficient to preserve positivity. 
\begin{figure}
\centering
	\includegraphics[width=1.0\textwidth]{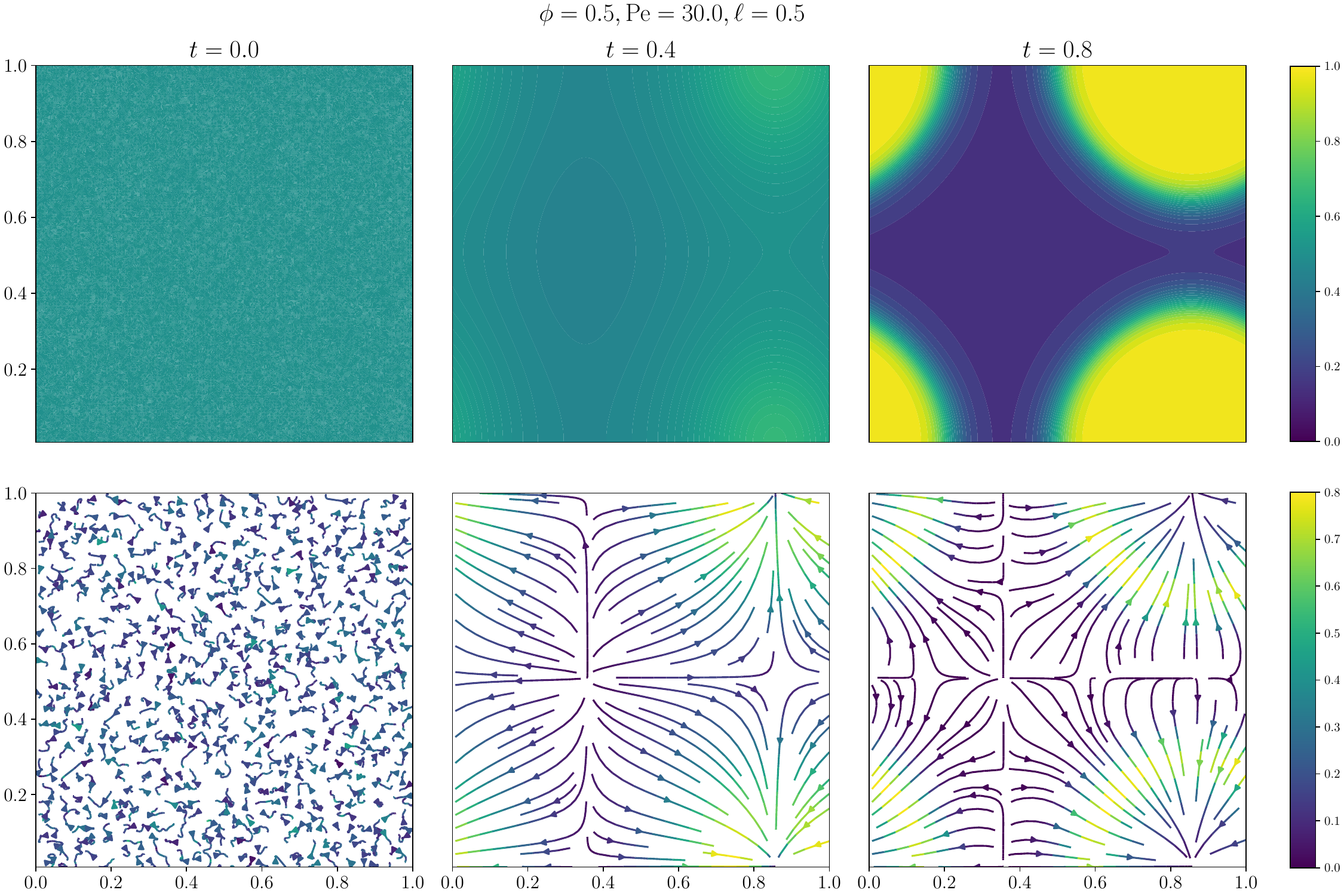} \hfill \vspace{0.01cm}
	\includegraphics[width=1.0\textwidth]{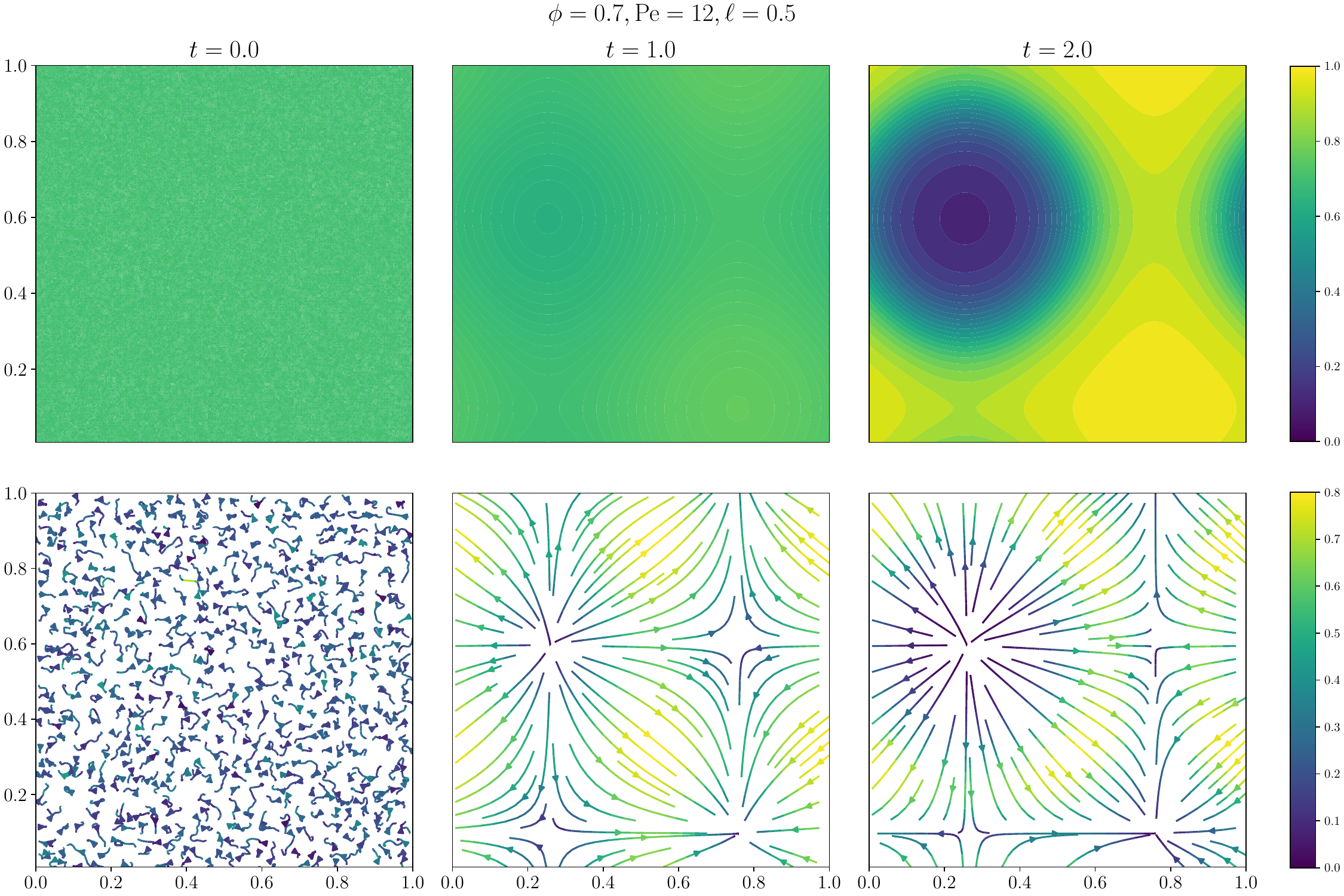}
	\caption{Example of two-dimensional patterns of the \emph{macroscopic} model \eqref{act_pas_eq} corresponding to $\phi= 0.5$, $\textnormal{Pe} = 30.0$ and $\phi= 0.7$, $\textnormal{Pe} = 12.0$ for $\Dx = 1.0$,  $\ell = 0.5$, $N_{x_1}= N_{x_2} = 128$, $N_\theta = 64$ starting from a random uniform perturbation of the uniform steady state.  Time advances from left to right.  The first and third rows correspond to the density $\rho(\x,t)$, and the second and fourth rows show stream plots of the polarisation $\mag(\x,t)$. 
	}
		\label{fig_3}
\end{figure}

Figure \ref{fig_3} shows two examples of instability from an initial random perturbation $f_0(x,\theta) \sim \textnormal{Unif}(\frac{\phi}{2\pi}-\delta,\frac{\phi}{2\pi}+\delta)$. The particle density $\rho$ is displayed in the first row, and the polarisation $\mag$ is displayed in the second row. In both cases,  the system phase separates into high and low-density regions. We observe that larger spatial gradients occur in regions with significant polarization. Roughly speaking, the polarisation tends to lie parallel with the density gradient, so that the diffusive and advective terms balance in \eqref{act_pas_eq}.
\begin{figure}
\centering
\includegraphics[width=.6\textwidth]{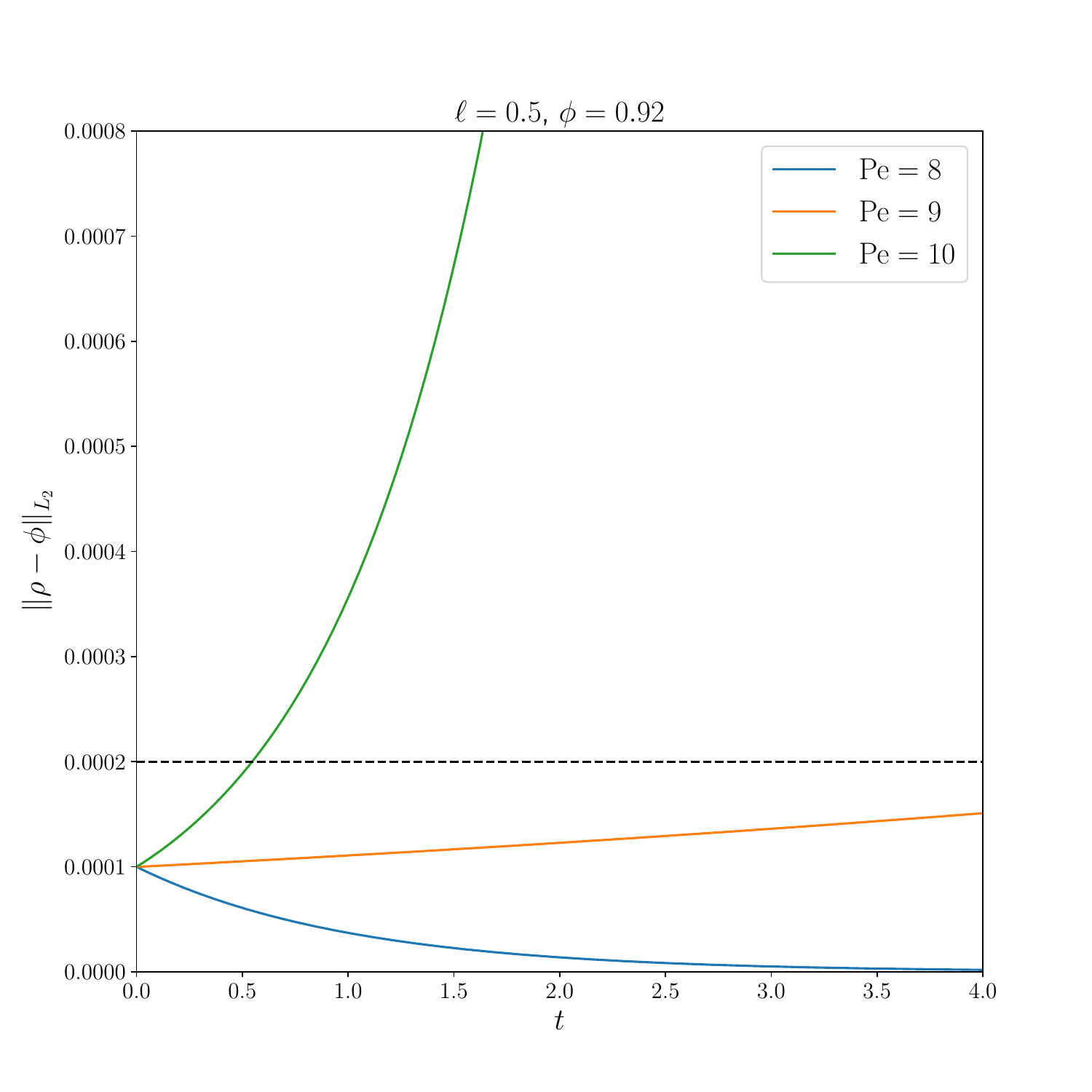}
	\caption{Evolution of the $\tw{L}^2$ distance \eqref{equ_discrete_H2} of the particle density $f$ from uniform, with initial condition $f_0$ given by \eqref{equ_discrete_H2}. For parameters $\Dx = 1.0$, $\phi = 0.92$, $\ell = 0.5$ with stable case, $\textnormal{Pe} = 8.0$, the unclassified case $\textnormal{Pe} = 9.0$ and unstable case $\textnormal{Pe} = 10.0$.
	} 
	\label{fig_4}
\end{figure}

\subsection{Nonlinear instability}
\label{sec_nonlinear}

To analyse the stability of the homogeneous solution to \eqref{act_pas_eq},
we solve the macroscopic model \eqref{act_pas_eq} using the finite volume scheme described in Subsection \ref{sec_pde_scheme} with an initial perturbation around the homogeneous state. In particular, we take the initial condition to be the eigenfunction corresponding to the eigenvalue with the largest positive real part, as given by \eqref{equ_pert}. That is %The initial condition is then 
\begin{equation}\label{equ_initial_pert}
	f_0(\x, \theta) = \frac{\phi}{2 \pi} + \delta \operatorname{Re}{\Bigg[ \sum_{0 \leq k \leq n} A^n_k \cos( k \theta) \exp( i \omega x_1 ) \Bigg]},
\end{equation}
where $\Re{x}$ represents the real part of $x$.
Having obtained a suitable numerical solution, we study the growth of the perturbation over time using a discretized $L^2$ norm:
%We discretize the $L^2$ norm to study the growth of the perturbation over time
\begin{align}\label{equ_discrete_H2} 
	\Vert g \Vert_{\tw{L}^2} &= \frac{1}{ \Delta {x_1}\Delta {x_2} \Delta \theta } \sum_{i,j,k} g_{i,j,k}^2.
\end{align}
Figure \ref{fig_4} shows illustrative results.  We take parameters $\delta = 10^{-4}$, $n = 40$, $\omega = 2 \pi$ and the diffusive lengthscale $\ell = 0.5$.  We vary Pe between $8$ and $10$ (as shown), and we
%Figure \ref{fig_1} shows four sample outputs for $\textnormal{Pe} = 6.0,8.0,8.2,8.4,10.0$. We 
classify each case as
\vspace{-6pt}\par  %% suppress the space before the start of the list
\begin{enumerate}[$\quad$-]
	\item unstable if $ \sup\limits_{0 \leq t \leq 4} \Vert f(t) - \frac{\phi}{2 \pi}\Vert_{\tw{L}^2} > 2 \delta$; 
	\item stable if $ \sup\limits_{0 \leq t \leq 4} \Vert f(t) - \frac{\phi}{2 \pi}\Vert_{\tw{L}^2} \leq 2 \delta$ and $ \partial_t \Vert f(t) - \frac{\phi}{2 \pi}\Vert_{\tw{L}^2} \leq 0$ at $t=4$;
	\item unclassified if $ \sup\limits_{0 \leq t \leq 4} \Vert f(t) - \frac{\phi}{2 \pi}\Vert_{\tw{L}^2} \leq 2 \delta$ and $ \partial_t \Vert f(t) - \frac{\phi}{2 \pi}\Vert_{\tw{L}^2} > 0$ at $t=4$.
\end{enumerate}

In particular $\textnormal{Pe} = 10.0$ is classified as unstable as it surpasses $2 \delta = 2 \times 10^{-4}$, indicated with a dashed line. The case $\textnormal{Pe} = 8.0$ is classified as stable as it is decreasing and below $2 \delta$, whereas $\textnormal{Pe} = 9.0$ is unclassified because it is below $2 \delta$ but increasing at the final time $t$.

To obtain Figure \ref{fig_1}, we iterated this process and performed a
%We next do a 
parameter sweep for $\phi = 0.44, 0.46, \dots, 1.0$ and $\textnormal{Pe} = 0, 1, \dots 40$.
%We plot the results in Figure \ref{fig_1}. 
For reference, we also plot the limit of linear stability \eqref{equ_matrix_system}. %derived in Section~\ref{sec_lin_stab}.
The numerically computed instability matches the limit of linear stability across all densities.

\subsection{Comparison of macroscopic and microscopic systems}

\begin{figure}
\centering
\includegraphics[width=1.0\textwidth]{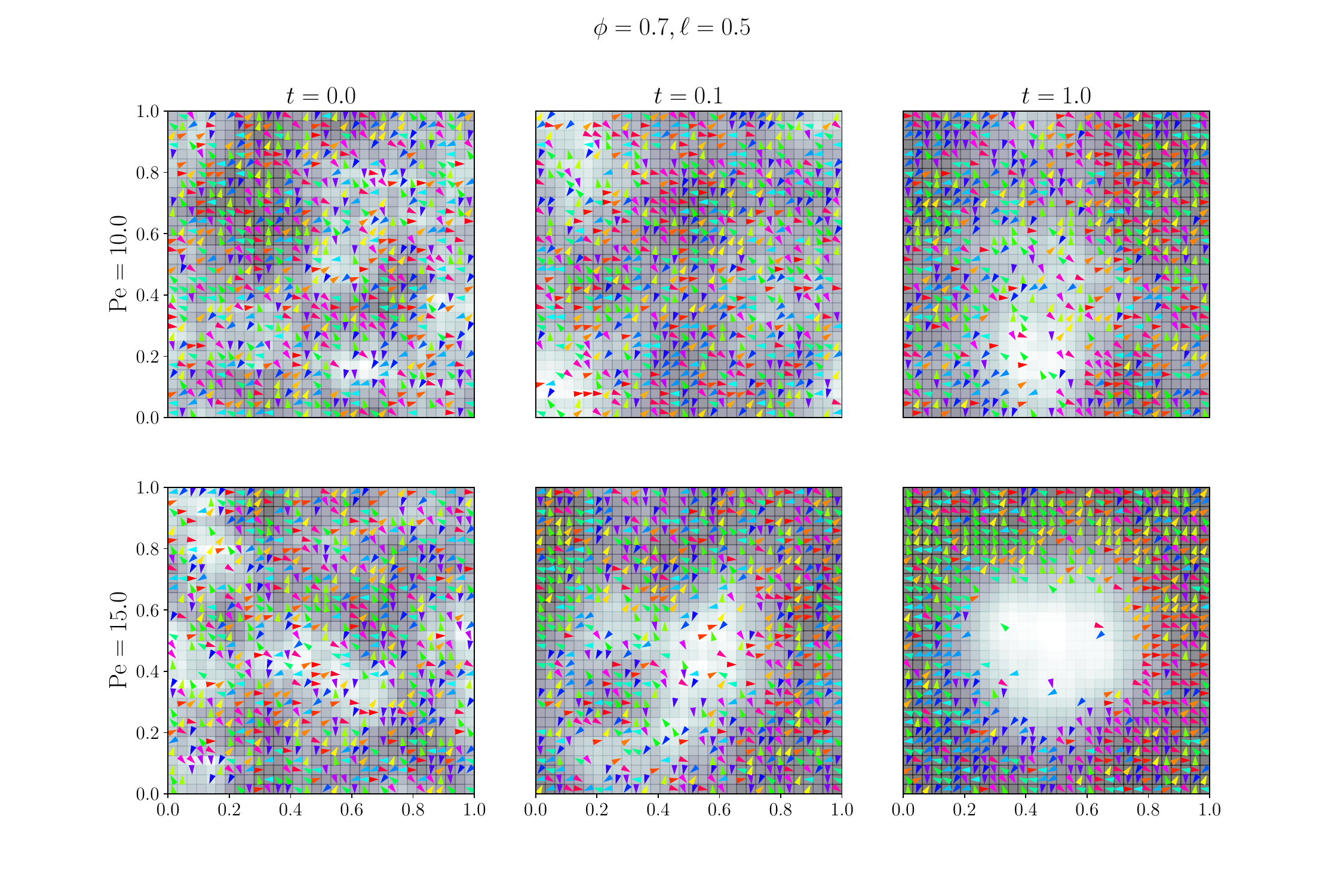}
	\caption{Snapshots of an output of the microscopic model at $T= 0.0,0.1,1.0$ %for $N= 32$, $\Dx = 1.0$, $\phi= 0.7$, $\textnormal{Pe} = 10.0, 15.0$, $\ell = 0.5$. 
	The particles are represented by triangles of size $\frac{1}{N}$ pointing towards $\e_{\theta_x}$ with colours corresponding to their orientations. Each cell is also shaded in proportion to the local particle density. %with $\ep = \frac{1}{16}$.
	}
	\label{fig_5}
\end{figure}

In this section, we compare the results obtained from the macroscopic equation with stochastic simulations of the microscopic models. For comparison with the macroscopic case, we define
the local particle density in the microscopic model:
\begin{equation}
	\rho^{\ep N}_{\x} ({\hat \eta} ) =  %\Big\langle \pi^N, \Big( \frac{1}{2\ep} \Big)^2 \id_{[-\ep,\ep]^2}  \Big(\cdot - \frac{x}{N} \Big)  \Big\rangle = 
	 \frac{1}{(2\ep N+1)^2} \sum_{\Vert N\x - {\z} \Vert_\infty \leq \ep N} \eta_{{\z}} ,
\end{equation}
%is plotted in the first row 
and the local polarisation is
\begin{equation}
	%\Big(\rho_{\ep N}^{\cos}\Big(\frac{z}{N},t \Big), \rho_{\ep N}^{\sin} \Big(\frac{z}{N},t \Big)\Big)  =  %\Big\langle \pi^N, \Big( \frac{1}{2\ep} \Big)^2 \id_{[-\ep,\ep]^2}  \Big(\cdot - \frac{x}{N} \Big)  \Big\rangle =
	\mag^{\ep N}_{\x} ( {\hat \eta} )=
	 \frac{1}{(2\ep N+1)^2} \sum_{\Vert N\x - {\z} \Vert_\infty \leq \ep N} \eta_{{\z}} (\cos ( \theta_{{\z}} ), \sin(\theta_{{\z}} ) ) .
\end{equation}

We initialise the model according to Section \ref{sec_micro_model}, with $N^2$ sites and a uniform initial profile ($\hat \zeta \equiv \frac{\phi}{2 \pi}$). The system dynamics combines both a jump and diffusion process.
A fixed timestep $\Delta t$ is used to discretize the diffusion process. 
Due to the separation of scales, the timestep can be chosen such that $ N^{-2} \Dx \ll \Delta t \ll 1 / \Dt$, we take $\Delta t=0.001$.
For each time increment $\Delta t$, we run a Gillespie algorithm, with fixed $\theta_{\z}$, until the time elapsed $\delta t$ surpasses $\Delta t$. The Gillespie algorithm simulates a finite state space Markov chain exactly \cite{erbanStochasticModellingReaction2020} . At each time step we first sample the next time interval according to an exponential distribution. We then sample a jump, weighted by each jumps probability of occurring first. We then execute the jump and update the corresponding jump rates. Once the time elapsed $\delta t$ surpasses $\Delta t$,  we update each $\theta_x$, sampling from ${\cal N} ( \theta_x, 2 \Dt \delta t)\mod 2\pi$. 

Figure \ref{fig_5} shows two examples of the system at $t = 0.0, 0.1, 1.0$ for $N=32$, $\phi = 0.7$ and $\textnormal{Pe} = 10,15$. The particles are represented by triangles of size $\frac{1}{N}$ pointing towards $\e_{\theta_x}$ with colours corresponding to their orientations. Each cell is also shaded in proportion to the local particle density, with $\ep = \frac{1}{16}$.

\begin{figure}
\centering
	\includegraphics[width=1.0\textwidth]{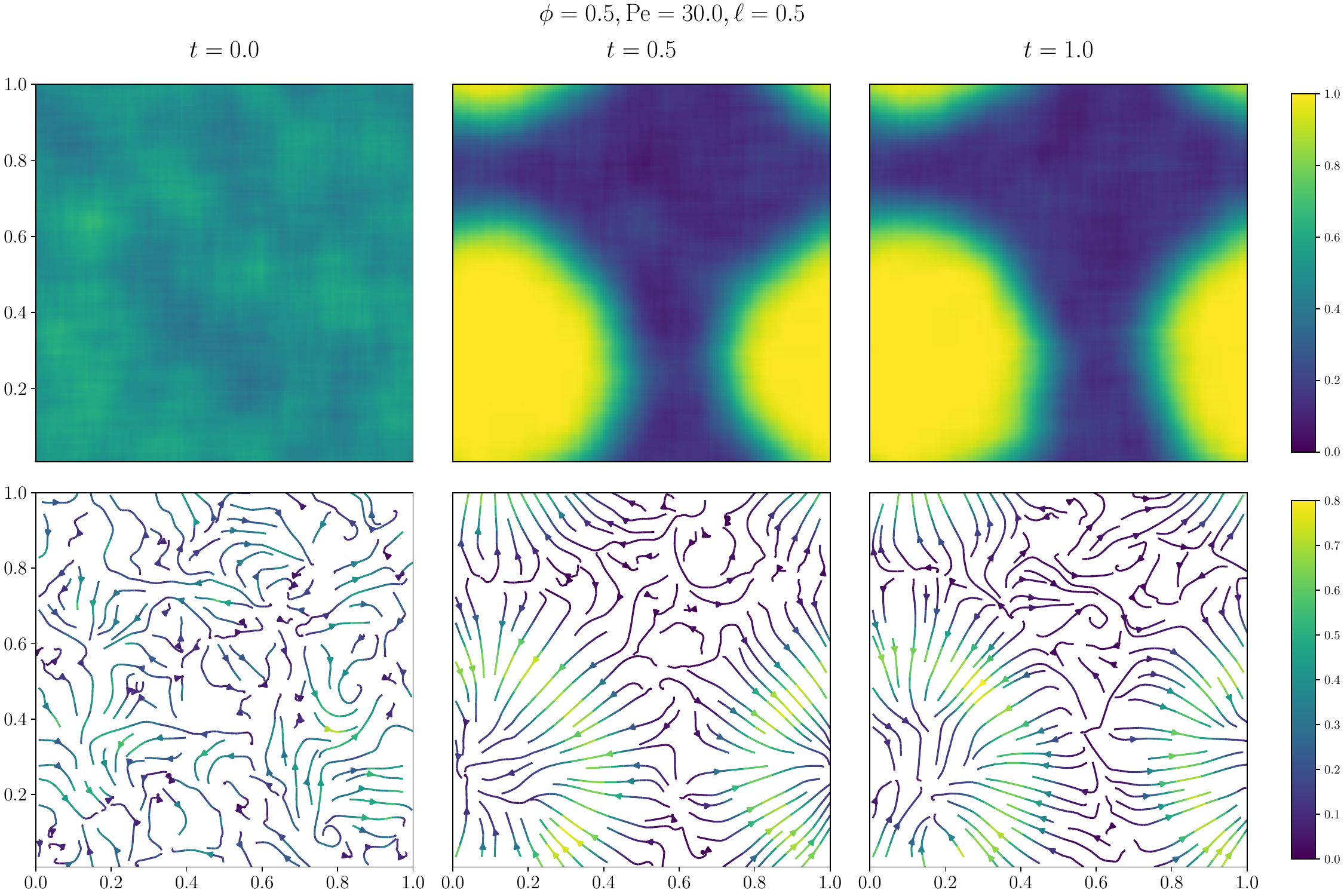} \hfill \vspace{0.01cm}
	\includegraphics[width=1.0\textwidth]{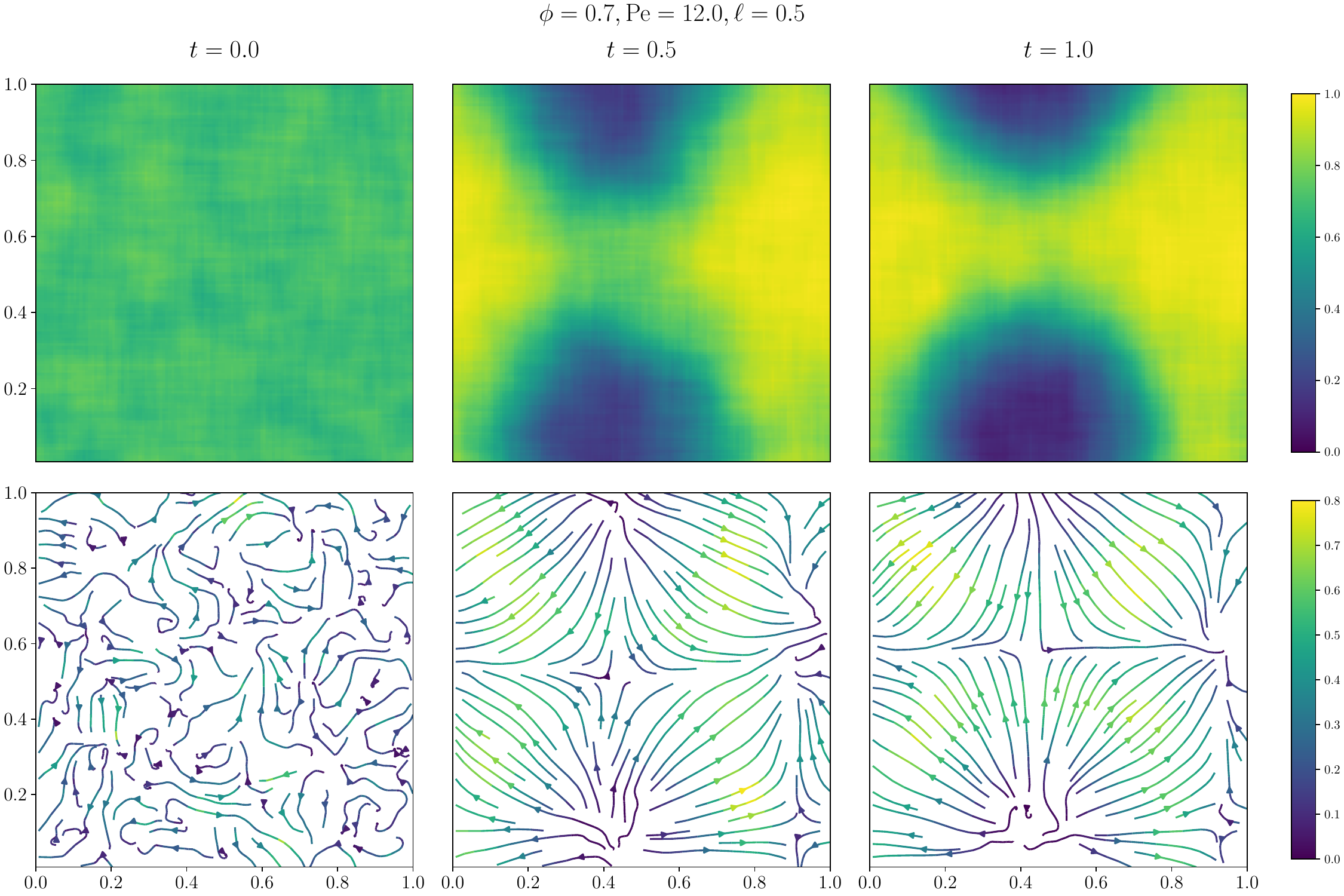}
	\caption{Example of a two-dimensional pattern of the \emph{microscopic} model for $N = 128$, corresponding to $\phi= 0.5$, $\textnormal{Pe} = 30.0$ and $\phi= 0.7$, $\textnormal{Pe} = 12.0$ for $\Dx = 1.0$,  $\ell = 0.5$, starting from the uniform distribution. The local particle density $\rho^{\ep N}$ is plotted in the first row, and the local polarisation is plotted in the second row, for $\ep = \frac{1}{32}$. In particular, we see good agreement with the patterns formed by the macroscopic model shown in Figure \ref{fig_1}.}
	\label{fig_6}
\end{figure}

Figure \ref{fig_6} shows two examples of the system at $t = 0.0, 0.5, 1.0$ for $N=128$ and $\phi = 0.5, \textnormal{Pe} = 30$ and $\phi = 0.7, \textnormal{Pe} = 12$. 
%is plotted in the second row. The quantities are averaged over a macroscopic radius of $\ep = \frac{1}{16}$. 
These results show that MIPS is present for both $\textnormal{Pe} = 12,30$.  The particles in each system self-organise into a dilute region with almost zero density and a dense region with particles in close packing. We also observe a strong particle alignment in the boundary between the two regions.

We emphasise that while the initialisation protocol of the system is homogeneous and invariant under spatial translations, dense and dilute regions have developed.  This is a spontaneous breaking of translational symmetry in the occupation.  This effect can be quantified by defining~\cite{yuPerpendicularParallelPhase2022,kornissNonequilibriumPhaseTransitions1997}
\begin{equation}\label{equ_trans_sym}
	\Phi({{{\hat \eta}}}) =  \Bigg\vert \frac{1}{N^2} \sum_{{\z} \in \lat} \eta_{\z} \exp \Big( \frac{2 \pi i z_1}{N} \Big)\Bigg\vert + \Bigg\vert \frac{1}{N^2} \sum_{z \in \lat} \eta_{\z} \exp \Big( \frac{2 \pi i z_2}{N} \Big)\Bigg\vert,
\end{equation}
which is
the sum of the magnitudes of the first Fourier modes in the horizontal and vertical direction. %to quantify the translational asymmetry in site occupation. 
In a homogeneous system then $\mathbb{E}[\Phi] = O(1/N)$,  but symmetry-broken states have $\mathbb{E}[\Phi]=O(1)$.

\begin{figure}
\centering
	\includegraphics[width=1.0\textwidth]{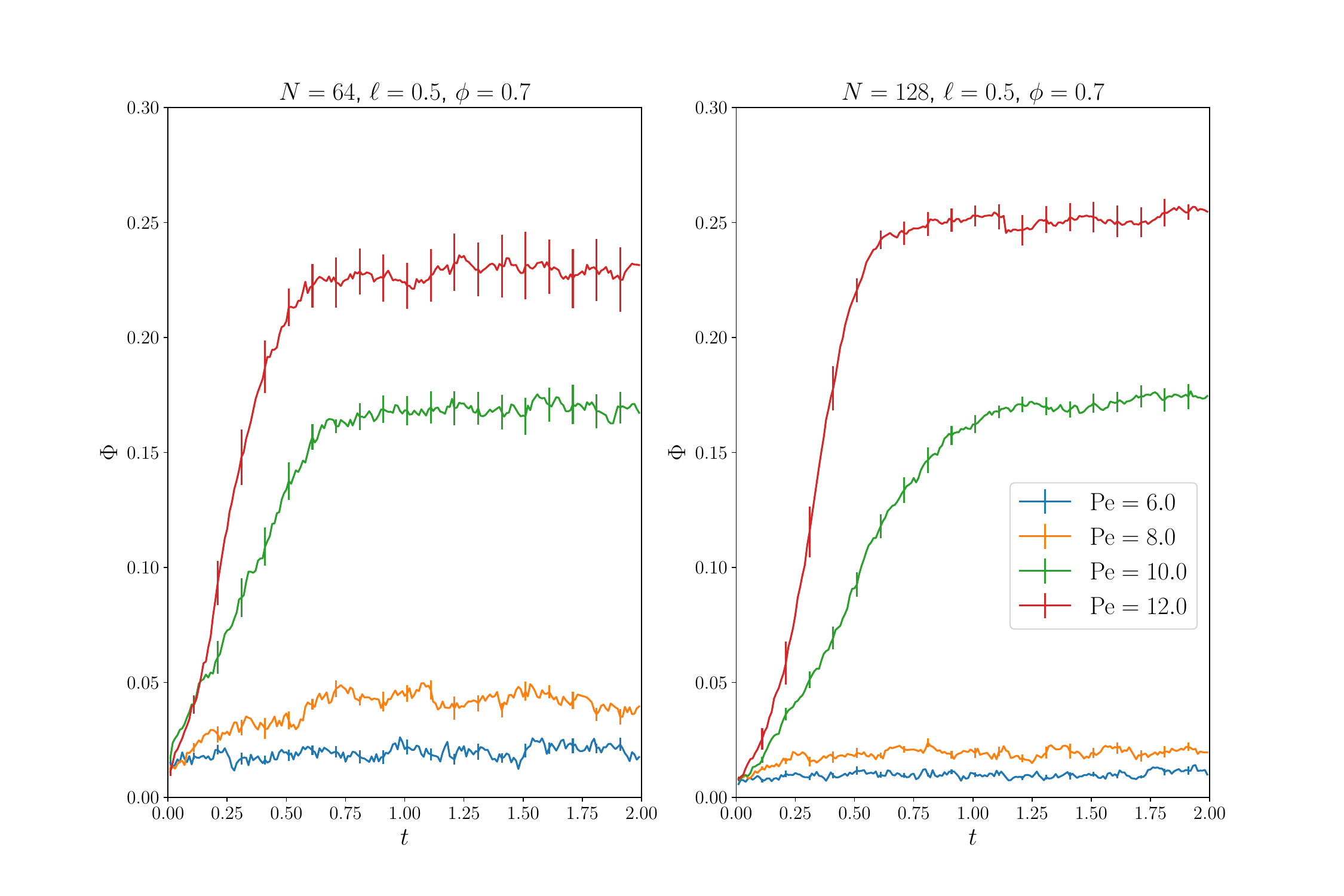}
	\caption{Evolution of the translational asymmetry $\Phi({{{\hat \eta}}}_t)$ given by \eqref{equ_trans_sym} of the microscopic model for $N= 64$ and $N= 128$. For unstable cases $\Dx = 1.0$, $\phi = 0.7$, $\textnormal{Pe} = 6.0, 8.0$, $\ell = 0.5$ and stable cases $\Dx = 1.0$, $\phi = 0.7$, $\textnormal{Pe} = 6.0, 8.0$, $\ell = 0.5$. Solid lines display the sample mean $\langle \Phi({{{\hat \eta}}}_t) \rangle$ averaged over ten samples, and vertical bars display the standard error. }
	\label{fig_7}
\end{figure}

Estimates of %this expectation value 
$\Phi({{{\hat \eta}}}_t)$, averaged over 10 samples, are shown in Figure \ref{fig_7} for $N = 64,128$ and $\textnormal{Pe} = 6,8,10,12$. They indicate that the system is macroscopically homogeneous for ${\rm Pe}=6,8$, but the symmetry is spontaneously broken for ${\rm Pe}=10,12$.
The stability of the macroscopic model displayed in Figure \ref{fig_1} is in good agreement with Figure \ref{fig_7}, for both which $\textnormal{Pe} = 6,8$ are stable and $\textnormal{Pe} = 10,12$ are unstable.

\begin{figure}
\centering
	\includegraphics[width=1.0\textwidth]{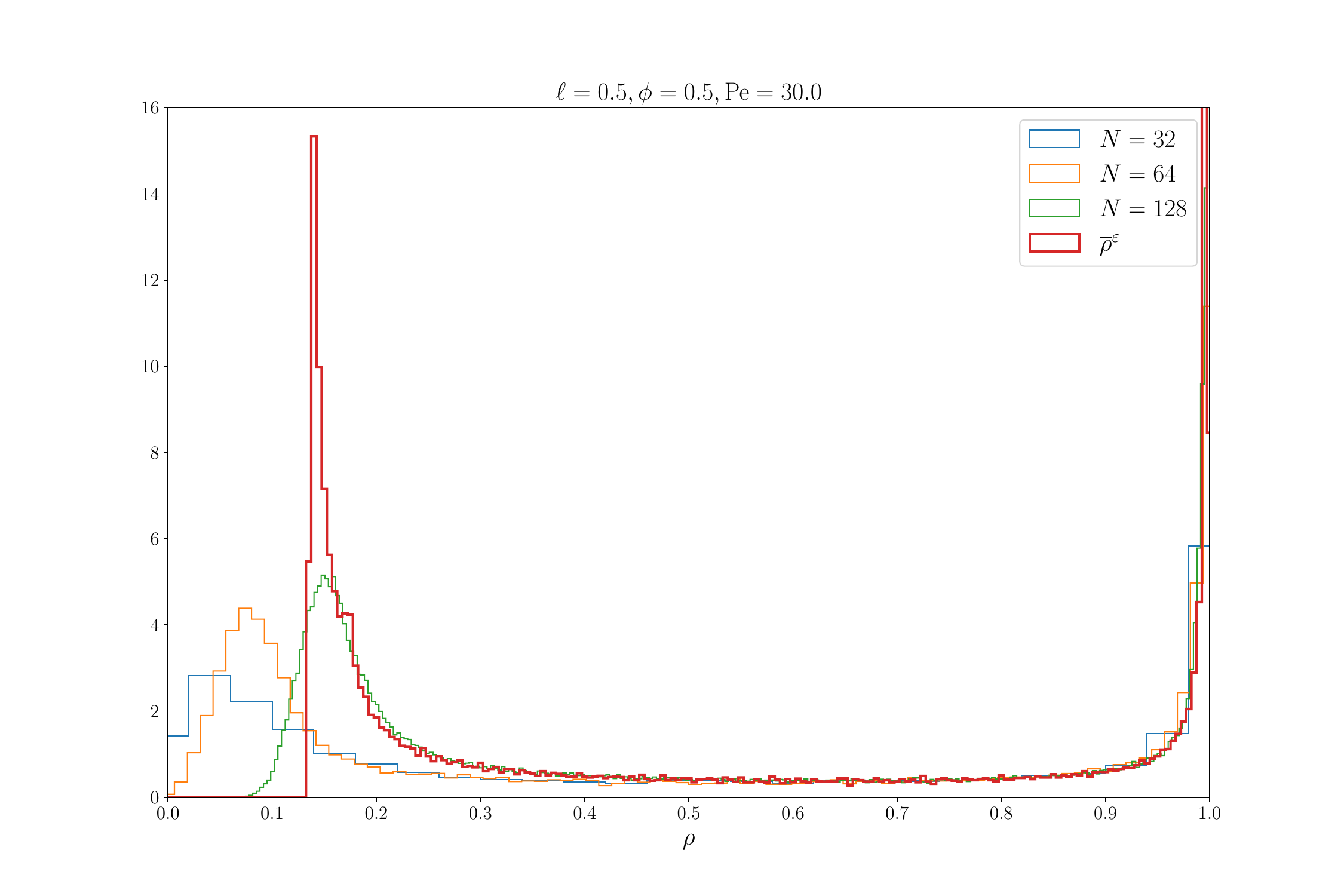}
	\caption{Histograms of the local density in the stationary state of the microscopic model $N = 32, 64, 128$ and the macroscopic model, with parameters $\Dx = 1.0$, $\phi = 0.5$, $\textnormal{Pe} = 30$, $\ell = 0.5$.}
	\label{fig_8}
\end{figure}

We close this Section with a quantitative comparison between the steady states of the microscopic and macroscopic models. To this end, we take parameters $\phi = 0.5, \textnormal{Pe} = 30, \ell = 0.5$.  For the microscopic model,
we compute the local density $\rho^{\ep N}$,
with $\ep = \frac{1}{16}$, for $N = 32, 64, 128$. After waiting for the system to reach equilibrium, we averaged over times $t = 2.0,2.1,\dots,4.0$.  Combining data from six such realisations, we plotted histograms of the local density, shown in Fig.~\ref{fig_8}.  Since the homogeneous steady state is unstable, the system phase separates, and two peaks emerge; the low-density peak corresponds to the dilute (``gaseous'') phase; the dense phase is a ``liquid''.
The local maxima of the histogram are the modal densities of the two phases.  

To compare this microscopic result with the macroscopic description, we define a corresponding local density
\begin{equation}
	{\bar \rho}^\ep(t,\x) = (2 \ep)^{-2}  \int_{\{\Vert \y-\x\Vert_\infty < \ep\}} \rho(\y,t) \dd y,
\end{equation}
which we compute from our numerical solutions of the hydrodynamic equation, after initialisation with a random perturbation around the uniform steady state, and averaged over times $t = 2.0,2.1 \dots, 4.0$. Histograms of this local density are also shown in Figure \ref{fig_8}.
For increasing $N$, the histograms computed from the microscopic model converge towards the result of the macroscopic model, as predicted by Theorem \ref{thm_main}. 

\section{Hydrodynamic Limit}
\label{sec_hydrolim}

In this section, we give a precise statement of the hydrodynamic limit formally described in \eqref{act_pas_eq}. To arrive at this macroscopic description of our \macc, we define its empirical measure.  The hydrodynamic limit is then obtained as a characterisation of this (random) measure in the limit $N\to\infty$.  
Let ${\cal M}(\TT^2 \times \SS)$ denote a non-negative measure on the continuous configuration space endowed with the weak topology and
\begin{equation}
	{\cal M}^{[0,T]} = D( [0,T], {\cal M}(\TT^2 \times \SS))
\end{equation}
denote the space of right continuous and left limited trajectories on ${\cal M}(\TT^2 \times \SS)$. 
Each trajectory of the process $\bconfig = ( {{{\hat \eta}}}(t) )_{t \in [0,T]}$ admits a natural image in ${\cal M}^{[0,T]}$ through its empirical measure
\begin{equation}
\label{eq:defEM}
	\pi^N_t  = \frac{1}{N^2}\sum_{{\z}\in \lat} \eta_{\z}(t) \delta_{\frac{{\z}}{N}, \theta_{\z}(t)} \in {\cal M}(\TT^2 \times \SS),
\end{equation}
where $\delta_{x,\theta}\in {\cal M}(\TT^2 \times \SS)$ is Dirac measure at $(x,\theta)$. Note that $\pi_t^N$ can equivalently be characterised by its integral against a test function $H$, see \eqref{eq:empmeasH}.
We endow ${\cal M}^{[0,T]}$ with Skorohod's metric, and the set ${\cal P} ( {\cal M}^{[0,T]} ) $ of probability measures on ${\cal M}^{[0,T]}$ with the weak topology. We now define 
\begin{equation}
\label{eq:DefQN}
Q^N \in {\cal P} ( {\cal M}^{[0,T]} )
\end{equation} 
as the distribution of the trajectory $\pi^N = ( \pi^N_t )_{t \in [0,T]}$ of the empirical measure of our process $\bconfig = ( {{{\hat \eta}}}(t) )_{t \in [0,T]}$. In deriving the hydrodynamic limit of this model, we explain that this trajectory $\pi^N_t(\dd x, \dd\theta)$ converges as $N\to\infty$, in probability, to a deterministic trajectory   $\pi_t(\dd \x, \dd\theta)=f(\x, \theta, t)\dd \x \dd \theta$, whose density $f$ represents the local density of particles with angle $\theta$, and is a solution of a  hydrodynamic equation \eqref{act_pas_eq}. Throughout this section, through abuse of notation, we will use a subscript to denote the time argument, $f_t(\x,\theta) = f(t,\x,\theta) $.

\begin{definition}[Weak solution to \eqref{act_pas_eq}]
\label{def_weak_sol}
We call a measure-valued trajectory  $(\pi_t)_{t \in [0,T]} \in {\cal M}^{[0,T]}$  a weak solution to \eqref{act_pas_eq}, with initial condition ${\hat \zeta}(\x, \theta)$, if the following are satisfied
\begin{enumerate}
	\item $\pi_0(\dd \x,\dd \theta) = {\hat \zeta}(\x, \theta)\dd x \dd \theta$.
	\item For all $t \in [0,T]$, the measure $\pi_t$ is absolutely continuous w.r.t the Lebesgue measure on $\TT^2\times \SS$, that is, there exists a density profile $f: [0,T] \times \TT^2\times \SS  \to [0,+\infty)$, such that $\pi_t(\dd \x,\dd \theta) = f_t (\x, \theta) \dd \x \dd\theta$.
	\item The function $ \rho_t(\x) = \int_\SS f_t(\x,\theta)\dd \theta $ is  in $L^2([0,T]; H^1(\TT^2) )$, and is therefore spatially differentiable.
	\item For all functions $H \in C^{1,2,2}([0,T] \times \TT^2 \times \SS)$,
	\begin{multline*}
		\langle f_T , H_T \rangle - \langle f_0 , H_0 \rangle  \\
        = \int_0^T  \langle f_t , \p_t H_t \rangle \dd t  + 
		\int_0^T \int_{\TT^2\times \SS} \Bigg\{ \Dx \Big[ \nabla^2 H \ds(\rho) f-\nabla H \cdot \nabla \rho (\vd(\rho) \\  - \ds^\prime(\rho) )   f  \Big]  	
        + v_0 \nabla H \cdot \Big[ \mag s(\rho) f+ {{{ \boldsymbol{e}_{\theta} }}} \ds(\rho) f 
		\Big] + \Dt \p^2_\theta H f 
		\Bigg\}  \dd t \dd x \dd \theta ,
	\end{multline*}
	where $\langle f , g\rangle = \int_{\TT^2\times \SS} f(\x,\theta) g( \x,  \theta) \dd \x \dd\theta$.
\end{enumerate}
\end{definition}
We can now state precisely what it means for \eqref{act_pas_eq} to be the hydrodynamic limit of our system.
\begin{theorem}[Hydrodynamic Limit]\label{thm_main}
	The sequence $(Q^N)_{N \in \NN}$ \eqref{eq:DefQN} is weakly relatively compact, and any of its limit points $Q^*$ are concentrated on trajectories $\pi \in {\cal M}^{[0,T]}$ which are weak solutions of \eqref{act_pas_eq} in the sense of Definition \ref{def_weak_sol}.
\end{theorem}
As previously mentioned, the rigorous derivation of the hydrodynamic limit stated in Theorem \ref{thm_main} for our ALG is extremely technical.
In the rest of this section, we sketch the main arguments to prove Theorem \ref{thm_main}, following the structure of \cite{erignouxHydrodynamicLimitActive2021}, stating the most crucial estimates explicitly and outlining the necessary adaptations due to the different angular dynamics considered here (angular diffusion, instead of an angular run-and-tumble type jump process in \cite{erignouxHydrodynamicLimitActive2021}.

\subsection{Dynkin formula and replacement Lemmas}

To make sense of the convergence from a discrete lattice to its continuum hydrodynamic limit, we therefore consider the \emph{empirical measure $\pi^N_t$ of the {\model}}, defined in \eqref{eq:defEM}. According to \eqref{eq:empmeasH}, its integral against test functions $H \in C^{1,2,2}([0,T] \times \TT^2 \times \SS)$ is 
\begin{equation}
\langle \pi^N_t , H_t \rangle:=\frac{1}{N^2}\sum_{{\z}\in \lat}\eta_{\z}(t) H\left(\frac{{\z}}{N},\theta_{\z}(t)\right),
\end{equation}
which readily applies to functions that do not depend on the angle $\theta$. Our goal is to prove that this empirical measure converges as $N\to\infty$ to a distribution $f(t,\theta,\x) d \x d\theta$, and that $f$ is a solution to the hydrodynamic equation \eqref{act_pas_eq}.

To do so, we start by defining
\begin{equation}\label{matingale}
	M^{H,N}_t = \langle \pi^N_t , H_t \rangle -\langle \pi^N_0 , H_0 \rangle - \int_0^t (\p_s +L_N) \langle \pi^N_s ,  H_s \rangle \dd s,
\end{equation}
which represents the fluctuation of the microscopic system around its hydrodynamic limit. By Dynkin's formula $M^{H,N}_t$ is a martingale. One straightforwardly shows that the $M^{H,N}_t$ vanishes as $N \to \infty$ in the scaling we chose between the different parts of our dynamics. Indeed, the quadratic variation, $[M^{H,N}]$, of $M^{H,N}$ can be explicitly computed by the identity
\begin{equation}
\label{eq:quadraticvariation}
	0\leq [M^{H,N}]_t 
	= \int_0^t 	L_N \langle \pi^N_s ,  H_s \rangle^2 -2 \langle \pi^N_s ,  H_s \rangle L_N\langle \pi^N_s ,  H_s \rangle   \dd s \leq \frac{C_H t}{N^2},
\end{equation}
where $C_H$ is a constant (dependent on $H$). Hence $M^{H,N}$ vanishes as $N\to\infty$. We do not detail this estimate; it can be found, e.g. in \cite[p. 127]{erignouxHydrodynamicLimitActive2021}. 

In what follows, it will be convenient to see the angular dependence of the empirical measure as a Dirac measure $\delta_{\theta_{\z}}(\dd \theta)$ integrating against an angle-dependent function $H({\z}/N,\theta)$. With this in mind, we now turn to the integral term in \eqref{matingale}, for which we write, using discrete summations by parts over the test function $H$
\begin{multline}
\label{eq:Dynkin2}
		\int_0^t (\p_s+L_N) \langle \pi^N_s ,  H_s \rangle \dd s = \\ 
		\frac{1}{N^2} \int_0^t \dd s \Bigg[ \int_{\SS}\sum_{{\z} ,i} 
		(N j^s_{{\z}, {\z}+e_i} + j^a_{{\z}, {\z}+\e_i})(s, d\theta) \p_{x_i,N} H_s \Big(\frac{{\z}}{N}, \theta\Big)  \\
		 + \sum_{\z} \eta_{\z}(s) \delta_{\theta_{\z}(s)}(\dd \theta) [\Dt\p_{\theta}^2 H_s+\p_{s} H_s ] \Big(\frac{{\z}}{N}, \theta\Big) \Bigg]\\
		= \frac{1}{N^2} \int_0^t \dd s \int_{\SS}\sum_{{\z} ,i} 
		(N j^s_{{\z}, {\z}+e_i} + j^a_{{\z}, {\z}+\e_i})(s, d\theta) \p_{x_i,N} H_s \Big(\frac{{\z}}{N}, \theta\Big)
		\\
		+\int_0^t\langle \pi^N_s ,  \Dt \p^2_\theta H_s+ \p_s H_s \rangle \dd s
\end{multline}
where  we defined for any  smooth function $H:\TT^2\to\RR$  
\begin{equation}
	\p_{x_i, N} H (\x) = N \Big[ H \Big(\x+\frac{\e_i}{N} \Big) -H \Big(\x \Big) \Big],
\end{equation}
for the microscopic approximation of $\p_{x_i}H_t$. The currents $j^s$, $j^a$ are defined as the angular distributions
\begin{equation}\label{equ_sym_current}
j^s_{{\z},{\z}+\e_i}({{{\hat \eta}}}, \dd \theta) = \Dx\left[  \delta_{\theta_{\z}}(\dd \theta)\eta_{\z}  (1 - \eta_{{\z}+\e_i}) -\delta_{\theta_{{\z}+\e_i}}(\dd \theta) \eta_{{\z}+\e_i}(1 - \eta_{\z})\right],
\end{equation}
which represents the symmetric current of particles with angle in $[\theta, \theta+\dd \theta]$  going through the edge $({\z},{\z}+\e_i)$, and 
\begin{equation}j^a_{{\z},{\z}+\e_i}({{{\hat \eta}}},\dd \theta) =  \lambda_i(\theta_{\z})\delta_{\theta_{\z}}(\dd \theta) \eta_{\z}  (1 - \eta_{{\z}+\e_i}) - \lambda_i(\theta_{{\z}+\e_i})\delta_{\theta_{z+\e_i}}(\dd \theta)\eta_{{\z}+\e_i} (1 - \eta_{\z}),\end{equation}
which represents the asymmetric current going through $({\z}, {\z}+\e_i)$, and 
\begin{equation}\lambda_i(\theta)=\vo \e_i \cdot{{{ \boldsymbol{e}_{\theta} }}} = \vo \e_i \cdot (\cos (\theta), \sin(\theta) )^T\end{equation}
is the strength of the driving asymmetry in direction $i$ for a particle with angle $\theta$. 

\medskip

Note that because of the angular diffusion, the last term in \eqref{eq:Dynkin2} appears directly as a function of the empirical measure $\pi_s^N$. Further note the extra factor $N$ multiplying the symmetric current $j^s$, which we must balance by a discrete gradient. To prove the hydrodynamic limit, one needs to close Equation \eqref{eq:Dynkin2} by replacing the two remaining microscopic quantities, $j^s$ and $j^a$, by functions of the empirical measure, which are their macroscopic counterparts. This will allow us to take the limit $N\to\infty$ in \eqref{matingale}. Although  $j^s$ and $j^a$ are distributions in $ \theta$, we will, by abuse of language, refer to them as \emph{local functions of the configuration}, since they only depend on the configuration through a finite number of sites. Our goal is to replace local functions $g_{\z}(\hat\eta, \dd \theta)$ at site ${\z}$ by their average value, which is a function $\Phi(\rho_{\x}^{\varepsilon N}, \hat{\rho}_{\x}^{\varepsilon N}(\dd\theta))$ of the mesoscopic state of the process, where
\begin{equation}
\label{eq:rhozep}
\rho_{\x}^{\varepsilon N}(\hat\eta):=\frac{1}{(2\varepsilon N+1)^2}\sum_{\Vert {\z}-{N\x}\Vert_\infty\leq \varepsilon N}\eta_{{\z}}
\end{equation}
is the empirical mesoscopic density around site $\x = {\z}/N \in \TT^2$, and 
\begin{equation}
\label{eq:thetazep}
\hat{\rho}_{\x}^{\varepsilon N}(\hat\eta,\dd\theta):=\frac{1}{(2\varepsilon N+1)^2}\sum_{\Vert {\z}-{N\x}\Vert_\infty \leq \varepsilon N}\eta_{{\z}}\delta_{\theta_{{\z}}}(\dd \theta)
\end{equation}
is the empirical distribution of the angles in a mesoscopic box around position $\x = {\z}/N \in \TT^2$ in the macroscopic box.

\medskip

This substitution, of local functions by their average value, typically stated as a \emph{Replacement Lemma}, holds as $N\to\infty$ \emph{and then} $\varepsilon \to 0$.  One can see, for fixed $\varepsilon$ and letting $N\to\infty$, that both $\rho_{\x}^{\varepsilon N}$ and $\hat{\rho}_{\x}^{\varepsilon N}$ can be well approximated by functions of the empirical measure: given a site ${\z}=N\x$, for example, one easily checks that 
\begin{equation}
\label{eq:empmeasrho}
\rho_{\x}^{\varepsilon N}(\hat\eta)=\frac{1}{\ep^2}\langle \pi^N, {\bf 1}_{[\x-\ep, \x+\ep]}\rangle+O(1/N),
\end{equation}
where ${\bf 1}_{[\x-\ep, \x+\ep]} = \{ \y \in \TT^2: \| \x - \y \|_\infty \leq \ep\}$ is a macroscopic box of radius $\ep$.
Further for any bounded measurable function $\Phi:\SS\to\RR$
\begin{equation}
\label{eq:empmeasrho2}
\int_\SS \Phi( \theta) \hat{\rho}_{\x}^{\varepsilon N}(\hat\eta,\dd\theta)
=\frac{1}{\ep^2}\langle \pi^N, {\bf 1}_{[\x-\ep, \x+\ep]}\Phi\rangle+O(1/N),
\end{equation}
where ${\bf 1}_{[\x-\ep, \x+\ep]}\Phi:(x', \theta)\mapsto {\bf 1}_{[\x-\ep, \x+\ep]}(x')\Phi(\theta)$.

\subsection{Grand-canonical states and local equilibrium}

A crucial element in the Replacement Lemma, which  amounts to a local law of large numbers for the process, is to identify the local distribution at any given point ${\z}$ of the configuration, namely the probability  
\begin{equation}\PP_t(\dd {\hat\eta}_{{\z}'},\; {\z}'\in \Lambda({\z}))\end{equation} 
of observing at time $t$ a local configuration $\{{\hat\eta}_{{\z}'},\;{\z}' \in \Lambda({\z})\}$ in a large microscopic box $\Lambda({\z})$ around ${\z}$. Because the symmetric displacement of particles happens at a very fast rate $N^2$, non-conserved quantities relax quickly to equilibrium, and one can expect that the local distribution of the process will be closely approximated by the stationary states of ${\cal L}_\textnormal{S}$, parametrised by its conserved quantities. 

\medskip

The only conserved quantities for ${\cal L}_\textnormal{S}$ are the total number of particles and their angles.
Hence we fix a distribution $\hat \rho$ on $\SS$ with total mass $\rho:=\int_\SS \hat{\rho}(\dd\theta)\in [0,1]$. One then defines the grand-canonical states $\mu^\star_{\Lambda, \hat{\rho}}$ in a finite box $\Lambda\subset \ZZ^2$ as a product distribution where each site of $\Lambda$ is occupied by a particle independently with probability $\rho$, and particles' angles are chosen independently with distribution $\hat{\rho}/\rho$ :
\begin{equation}
\label{eq:mustar}
\mu^\star_{\Lambda, \hat{\rho}}(\dd {{{\hat \eta}}}):=\prod_{{\z}\in \Lambda}\left[(1-\eta_{\z})(1-\rho)+\eta_{\z}  \hat{\rho}( \dd \theta_{\z})\right].
\end{equation}

We denote by $\EE^\star_{\Lambda,\hat{\rho}}$ the corresponding expectation. 

\medskip

For the reasons listed above, we expect that the distribution of our Markov process at site ${\z}$ and macroscopic time $t$ is locally close, as $N\to\infty$ and then $\ep\to 0$, to 
$\mu^\star_{\Lambda, \hat{\rho}_{\z}^{\ep N}}$. More precisely, given a local function $g$ of the configuration, we expect that for any $\Lambda$ containing the support of $g$,
\begin{equation}
\label{eq:loceq}
\EE_N(g({{{\hat \eta}}}_{{\z}+\cdot}))=\EE^\star_{\Lambda,\hat{\rho}_{\z}^{\ep N}}(g)+o_{N\to\infty,\ep\to 0}(1).
\end{equation}
This property, called \emph{local equilibrium}, is a crucial argument in the proof of hydrodynamic limits.

\medskip

Fix a density profile $\act:\TT^2\times \SS \to\RR^+$, satisfying $\rho(\x):=\int_{\SS} \act(\x,\theta)\dd\theta\in [0,1]$, $\forall \x\in \TT^2.$  We associate to $\act$ a local equilibrium distribution,  defined analogously to \eqref{eq:mustar}
\begin{equation}
\label{eq:mustarf}
\mu^\star_{N,\act}(\dd {\hat \eta})=\prod_{{\z}\in \TT_N^2}[(1-\eta_{\z})(1-\rho({\z}/N))+\eta_{\z} \act({\z}/N,\theta_{\z})\dd \theta_{\z} ]
\end{equation}
This means that under $\mu^\star_{N,\act}$ each site $\z$ is occupied w.p. $\rho({\z}/N)$, and if it is occupied, the angle of the particle at site $z$ is distributed as $\act({\z}/N,\theta) \dd \theta/\rho({\z}/N)$; this is similar to our prescription for initialisation of the \macc.
%In other words, in the same way we chose the initial distribution of the {\macc},. 
Given a reference density $\rho^\star\in [0,1]$, we simply denote by $\nu_{\rho^\star}:=\mu^\star_{N,{\act}_{\rho^\star}}$ the equilibrium distribution associated with the uniform profile ${\act}_{\rho^\star}(\x,\theta)\equiv \frac{\rho^\star}{2\pi}$, in which each site is occupied with constant probability $\rho$, and particle's angles are chosen uniformly in $\SS$.

\begin{remark}[Mobility and Einstein's relation]
\label{rem:Einstein}
Shorten $f(\theta) :=f(\cdot,\cdot,\theta)$, and denote by $\delta_\theta(\dd \theta')$ the Dirac distribution at $\theta$ on $\SS$.  Recalling \eqref{eq:sigma1}, we introduce two parametric distributions
\begin{align}
\label{eq:Dandsigma}
D_{f,\theta}(\dd \theta')&=d_s(\rho)\delta_{\theta}(\dd\theta')+f(\theta) \vd (\rho ) f({\theta'}) \dd\theta',\notag \\ \sigma_{f,\theta}(\dd\theta')&= s(\rho ) f(\theta) f({\theta'}) \dd \theta' + \ds(\rho) f(\theta)\delta_{\theta}(\dd \theta'),
\end{align}
which respectively represent the cross-diffusion coefficient and the cross-mobility coefficient. We now introduce 
\begin{multline}
	\label{eq:distchi}
\chi_{f,\theta}(\dd\theta')=\sum_{\z\in \ZZ^2} \frac{\dd}{\dd \theta}\mathrm{Cov}_f(\eta_0{\bf 1}_{\{\theta_0\in \dd\theta\}},\eta_{\z}{\bf 1}_{\{\theta_{\z}\in \dd\theta'\}})\\
= f(\theta) \, \delta_{\theta}(d\theta')-f(\theta) f({\theta'})\dd\theta',
\end{multline}
which can be interpreted as cross-compressibility. Above, $\mathrm{Cov}_f(\cdot,\cdot)$ represents the covariance of the two variables with respect to the equilibrium state \eqref{eq:mustarf} on $\Lambda=\ZZ^2$. Since the latter is a product distribution, all contributions in \eqref{eq:distchi} vanish, except for $\z=0$, which gives the second identity in \eqref{eq:distchi}. 
%\medskip
Given a function $f(t,\x,\theta)$ and a distribution $\mu$ on $\SS$, define 
\begin{equation}
\langle \mu ,f\rangle_{\SS}:(t,\x)\mapsto \int_\SS  f(t,\x,\theta) \mu(\dd\theta),
\end{equation}
and the vector $\mathfrak{e}:\theta\mapsto \e_\theta$.
The hydrodynamic equation \eqref{act_pas_eq} straightforwardly rewrites 
\begin{equation}
\label{eq:hydroEinstein}
		\pt \act_\theta = \Dx \div \langle D_{f,\theta}, \nabla f \rangle_{\SS} - \vo \div \langle \sigma_{f,\theta}, \mathfrak{e}\rangle_{\SS}  + \Dt \p_\theta^2 \act_\theta. 
\end{equation}
Then, straightforward computations show that  the Einstein relation holds, namely 
\begin{equation}
\label{eq:Einsteineq}
\sigma_{f,\theta}(\dd\theta')= \langle D_{f,\theta},  \chi_{f,\cdot}(\dd \theta')\rangle_{\SS}:=\int_\SS \chi_{f,\theta''}(\dd \theta')  D_{f,\theta}(\dd \theta'').
\end{equation}
\end{remark}

\begin{remark}[Entropy and gradient-flow structure]
\label{rem:GradFlow}
We now define the relative entropy
\begin{equation}
H\big(\mu^\star_{N,{\act}}\,\big\vert\, \nu_{\rho^\star}\big):=\EE_{\nu_{\rho^\star}}\left(\frac{\dd \mu^\star_{N,{\act}}}{\dd\nu_{\rho^\star}}\log \frac{\dd \mu^\star_{N,{\act}}}{\dd \nu_{\rho^\star}}\right).
\end{equation}
Assuming that $\act$ is smooth, it is straightforward to show, by the law of large numbers, that the relative entropy, once suitably rescaled in $N$, is related to the free energy \eqref{equ:free-energy} as
%converges to the entropy functional 
\begin{equation}
%S[\act]:=S
\lim_{N\to\infty} \frac{H\big(\mu^\star_{N,{\act}}\,\big\vert\, \nu_{\rho^\star}\big)}{N}=
S[f] + C_1
%\int_{\TT^2} \left\{ (1-\rho(x))\log\frac{1-\rho(x)}{1-\rho^\star}+\int_{\SS} \act(x, \theta)\log\frac{2\pi \act(x, \theta)}{\rho^\star} \dd\theta \right\} \dd x
\end{equation}
for a suitable constant $C_1=C_1(\rho^\star)$. 
This relative entropy can be interpreted as the large deviation rate functional for the empirical measure $\pi^N$ in the {\macc}'s stationary state $\nu_{\rho^\star}$.
%\medskip
The functional derivative of $S$ is given by
%, for some constant $C=C(\rho^\star)$ by
\begin{equation}
\frac{\delta S}{\delta \act(\x,\theta)}=\log\frac{2\pi \act(\x,\theta)}{1-\rho(\x)}+C,
\end{equation}
which yields the factorised form of the hydrodynamic equation given in \eqref{eq:GFv0}.
\end{remark}

\subsection{Non-gradient replacement Lemma}
\label{sec:NGRL}

Note that the gradient $\p_{x_i,N} H_s $ in \eqref{eq:Dynkin2} comes from the difference of particle currents between current going in (from ${\z}-\e_i$ to $ {\z}$) and currents going out (from ${\z}$ to ${\z}+\e_i$) at any site ${\z}$, and balances out a factor $N$ in the symmetric and asymmetric parts of the generator $L_N$ (cf. Equation \eqref{equ_markov_gen}). By ``balancing out'', we mean that the extra factors $N$ rescaling the exclusion processes' generator (both symmetric and asymmetric parts) get absorbed into the discrete derivative of the test function so that the limiting contribution is of order $1$. One main difficulty in deriving the hydrodynamic limit of the {\macc} comes from the non-gradient nature of the generator, meaning that the symmetric part ${\cal L}_\textnormal{S}$ of $L_N$ does not readily act as a local Laplacian. More explicitly, the definition of gradient type is given in Remark 2.4 \cite{kipnisScalingLimitsInteracting1998} as 
 \begin{definition}[gradient-type] \label{def_grad_type}
Let $E$ be a subset of $\NN$, corresponding to particle occupation number. A translation invariant nearest neighbour particle system $\eta(t)$ on $E^{\TT^d_N}$ with generator $L_N$ is said to be gradient type if there exist cylinder functions $g_{i, n}$ and finite range functions $p_{i,n}$ such that the current can be written
	\begin{equation}
		j_{{\z},{\z}+\e_i}(\eta)= \sum_{n=1}^{n_0} \sum_{x \in \TT^d_N} p_{i,n}(x) \tau_{x} g_{i, n}(\eta),
	\end{equation}
 where $\tau$ is the translation operator and
 \begin{equation}
     \sum_{x \in \TT^d_N} p_{i,n}(x) = 0. 
 \end{equation}
\end{definition}
In our case, recalling that particles have orientations, this entails that the symmetric current $j^s_{{\z},{\z}+\e_i}$ \eqref{equ_sym_current} cannot be written in form of Definition \ref{def_grad_type}
\begin{equation}
\label{eq:gradfick}
j^s_{{\z},{\z}+\e_i}({{{\hat \eta}}}, \dd \theta)= \tau_{\e_i} g_{{\z}}({{{\hat \eta}}}, \dd \theta)- g_{{\z}}({{{\hat \eta}}}, \dd \theta),
\end{equation}
for any local function $g$ of the configuration. Note that the gradient condition \eqref{eq:gradfick} can be interpreted as a microscopic Fick's law and bears no relation to the gradient-flow form explored in Remark \ref{rem:GradFlow}. Without a gradient decomposition \eqref{eq:gradfick}, one cannot perform a second summation by parts on the test function, which would allow us to balance out the second factor $N$ appearing in the symmetric current.

As a result, we need to prove that $Nj^s_{{\z},{\z}+\e_i}$ can be replaced by a discrete gradient, which is also a function of the empirical measure. This replacement is much more complicated than for gradient systems, for which local equilibrium \eqref{eq:loceq} is enough, because instead of replacing a function of order $1$ by its average, we need to do so for a function of order $N$, meaning that the correction term in \eqref{eq:loceq} no longer vanishes.
In other words, non-gradient local equilibrium states are distorted so that we cannot replace $j^s$ by its average under the grand-canonical distribution, and lower order correlations to local equilibrium need to be taken into account in \eqref{eq:loceq}. Furthermore, although the asymmetric current $j^a$ is expected to be replaced by its local equilibrium average, lower-order corrections to local equilibrium \eqref{eq:loceq} need to be considered. This is a manifestation of Einstein's relation (cf. Remark \ref{rem:Einstein} above).
xIn particular, the asymmetric contribution in the hydrodynamic limit appearing in \eqref{act_pas_eq} is not simply the gradient of $\EE^\star_{\Lambda,fd\theta}(j^a_{0,e_i})$. More precisely, we have the following result:

\begin{proposition}[Non-gradient Replacement Lemma]
\label{lem_replace_II}
	We introduce 
	\begin{multline}
		Y_{{\z},i}^{\ep,N}({{{\hat \eta}}}, d\theta) = 
		N \bigg[ j^s_{{\z},{\z}+\e_i}+
		 \ds(\rho_{\x}^{\ep N})\left(\hat \rho_{{\x}+\e_i/N}^{\varepsilon N}
		-\hat \rho_{\x}^{\varepsilon N}\right)\\
  + \hat \rho_{\x}^{\varepsilon N}\vd (\rho_{\x}^{\varepsilon N} )  \left(\rho_{{\x}+\e_i/N}^{\varepsilon N}-\rho_{\x}^{\varepsilon N}\right) \bigg](\hat\eta,\dd \theta),
	\end{multline}
as well as 
		\begin{multline}
				Z_{{\z},i}^{\ep,N}({{{\hat \eta}}}, d\theta) 
				= j^a_{{\z},{\z}+\e_i}({{{\hat \eta}}},\dd \theta)
			\\	+  \hat \rho_{\x}^{\varepsilon N}({{{\hat \eta}}},\dd \theta) \Big[\lambda_i(\theta) \ds(\rho_{\x}^{\ep N}) 
				+ s(\rho_{\x}^{\ep N}) \int_\SS \lambda_i \dd \hat{\rho}_{\x}^{\ep N}\Big](\hat \eta),
		\end{multline}
	where $\x = \z/N \in \TT^2$.
	Then for any $H \in C^1 ([0,T] \times \TT^2\times \SS )$,  and $i \in \{1,2\}$,
	\begin{equation}
		\limsup_{\ep \to 0}\limsup_{N \to \infty} \EE_N \Bigg[ \Bigg\vert
		\int_0^T \frac{1}{N^2} \sum_{z \in \lat} \int_\SS H_t \Big (\x,  \theta \Big) \left[Y_{{\z},i}^{\ep,N} +Z_{{\z},i}^{\ep,N}\right]\dd\theta \dd t 
		\Bigg\vert  \Bigg] = 0.
	\end{equation}
\end{proposition}
Note that both  $Y_{{\z},i}^{\ep,N}({{{\hat \eta}}}, d\theta)$ and $Z_{{\z},i}^{\ep,N}({{{\hat \eta}}}, d\theta)$ above are differences between microscopic currents and their mesoscopic averages, and that the latter are functions of the empirical measure according to \eqref{eq:empmeasrho} and \eqref{eq:empmeasrho2}. Further note that we recovered in $Y_{{\z},i}^{\ep,N}({{{\hat \eta}}}, d\theta)$ microscopic gradients 
\begin{equation}\hat \rho_{{\x}+\e_i/N}^{\varepsilon N}-\hat \rho_{\x}^{\varepsilon N}\quad \mbox{ and }\quad \rho_{\x+\e_i}^{\varepsilon N}-\rho_{\x}^{\varepsilon N},\end{equation}
which will ultimately balance out the last factor $N$ by summation by parts in \eqref{eq:Dynkin2}.  The proof of Proposition \ref{lem_replace_II} is very technical, we do not reproduce it here and refer the interested reader to \cite[p. 128, Cor. 6.7.3 and Lem. 6.7.4]{erignouxHydrodynamicLimitActive2021} for a detailed implementation in the same setting.

\subsection{Proof of the hydrodynamic limit}

We are now in a position to close Equation \eqref{eq:Dynkin2} and take the limit $N\to\infty$. Recall that according to \eqref{eq:quadraticvariation} and \eqref{eq:Dynkin2},
\begin{multline}
\label{eq:Dynkin3}
		 \langle \pi^N_t , H_t \rangle- \langle \pi^N_0 , H_0 \rangle		+ \langle \pi_t^N [\p_{\theta}^2 H_t+\p_{t} H_t ] \rangle\\
		 - \frac{1}{N^2} \int_0^T \dd t \int_{\SS} \sum_{{\z} ,i} 
		(N j^s_{{\z}, {\z}+\e_i} + j^a_{{\z}, {\z}+\e_i})(t, d\theta) \p_{x_i,N} H_t (\x, \theta )
\end{multline}
 vanishes in probability as $N\to\infty$. Thanks to Proposition \ref{lem_replace_II}, the total current $N j^s_{{\z}, {\z}+\e_i} + j^a_{{\z}, {\z}+\e_i}$ in the last term can be replaced by
 \begin{multline}
 \label{eq:replacementNG} 
N\left[\ds(\rho_{\x}^{\ep N})\left(\hat \rho_{\x+\e_i}^{\varepsilon N}-\hat \rho_{\x}^{\varepsilon N}\right) + \hat \rho_{\x}^{\varepsilon N}\vd (\rho_{\x}^{\varepsilon N} )  \left(\rho_{{\x}+\e_i}^{\varepsilon N}-\rho_{\x}^{\varepsilon N}\right) \right](\hat\eta,\dd \theta)+\\  
\hat \rho_{\x}^{\varepsilon N}({{{\hat \eta}}},\dd \theta) \left[\lambda_i(\theta) \ds(\rho_{\x}^{\ep N}) + s(\rho_{\x}^{\ep N}) \int_\SS \lambda_i \dd \hat{\rho}_{\x}^{\ep N}\right](\hat \eta),
\end{multline}
where $\x = \z/N \in \TT^2$.
To consider the limit $N\to\infty$, consider the distribution $Q^N$ of the empirical measure's trajectory $(\pi^N_t)_{t\leq T}$, defined in \eqref{eq:DefQN}. We have the following result.

\begin{proposition}[Relative compactness and regularity of the density]
\label{prop:relcomp}
	The sequence $(Q^N)_{N\in\NN}$ is relatively compact, and any of its limit points $Q^*$ is concentrated on trajectories that are 
	\begin{enumerate}[i)]
	\item absolutely continuous with respect to the Lebesgue measure on $\TT^2\times \SS$, i.e. such that there exists a function $f$ such that $\pi_t( \dd \x , \dd \theta ) = f_t( \x ,\theta) \dd \x \dd\theta$. 
	\item such that the local density	
 \begin{equation}\rho_t(\x):=\int_\SS f_t(\x,\theta) d\theta\end{equation} 
	is in $  H^1([0, T ]\times \TT^2)$, meaning that for $i=1,2$, there exists two functions $\p_{x_i}\rho$ in $L^2([0,T]\times \TT^2)$,  such that for any smooth test function $H\in C^{0,1}([0,T]\times \TT^2)$, 
	\begin{equation}\int_{0}^T\int_{\TT^2}\rho_t(\x)\p_{x_i} H_t(\x) dt d\x=\int_{0}^T\int_{\TT^2}H_t(\x) \p_{x_i}\rho_t(\x)  dt d\x.\end{equation}
	\end{enumerate}
\end{proposition}

Note that this result says nothing, a priori, about the spatial regularity of $f$ itself, so that in the limit $N\to\infty$, there is no reason for $\hat \rho_{\x+\e_i/N}^{\varepsilon N}-\hat \rho_{\x}^{\varepsilon N}$ to converge to a well-defined quantity (in this case, $\partial_{x_i}f(\x,\theta)\dd\theta$. This is not a problem however, because $\rho$ admits spatial derivatives, and the self diffusion coefficient $d_s$ is $C^\infty$ \cite{landimSymmetricSimpleExclusion2001}, so that the discrete gradient $\hat \rho_{\x+\e_i/N}^{\varepsilon N}-\hat \rho_{\x}^{\varepsilon N}$ can be transferred  onto $d_s( \rho_{\x}^{\varepsilon N})$ and to the test function, as in Definition \ref{def_weak_sol}. 

\bigskip

In the limit $N\to\infty$, according to the previous proposition, for $\z=N\x$ we can replace $\rho_{\x}^{\varepsilon N}(\hat\eta(t))$ by $\frac{1}{\ep^2}\langle \pi^N, {\bf 1}_{[\x-\ep, \x+\ep]}\rangle$, which in turn can be replaced in the limit $\ep \to 0$ by $\rho_t(\x)$.  Similarly, $\hat{\rho}_{\x}^{\varepsilon N}(\hat\eta(t), \dd\theta)$ can be replaced as $N\to\infty$ then $\ep \to 0$ by $f_t(x,\theta) \dd\theta$. Finally, spatial averages can be replaced by integrals and discrete spatial derivatives by their continuous counterparts. This, together with the fact that \eqref{eq:Dynkin3} vanishes in probability and  by replacement  \eqref{eq:replacementNG}, proves that any limit point $Q^*$ of $(Q_N)$ is concentrated on trajectories $\pi$ satisfying i) and ii) above, and such that 
\begin{multline}
\label{eq:Dynkin4}
		 \langle f_t , H_t \rangle- \langle f_0 , H_0 \rangle		+ \langle f_t,[\p_{\theta}^2 H_t+\p_{t} H_t ] \rangle\\
		 +  \int_0^T \dd t \int_{\TT^2\times \SS}\dd x \dd\theta \sum_{i=1}^2  \Bigg[
	\left[\ds'(\rho_t) \p_{x_i} \rho_t \p_{x_i} H_t + \p_{x_i}^2 H_t\right]f_t+\\  
-\left[ f_t\vd (\rho_t )  \p_{x_i}\rho_t+f_t \lambda_i(\theta) \ds(\rho_t) + f_ts(\rho_t) \int_\SS \lambda_i(\theta)f_t(\theta) d\theta\right] \p_{x_i} H_t (\x, \theta)\Bigg]
\end{multline}
vanishes in $Q^*$-probability. Note we abuse our previous notation, in that we also denote by $\langle \cdot,\cdot \rangle$ the inner product in $L^2(\TT^2\times \SS)$. This proves Theorem \ref{thm_main}.

\begin{remark}
\label{rem:angulardiffusion}
In Proposition \ref{prop:relcomp} above, we slightly expanded on results in \cite{erignouxHydrodynamicLimitActive2021}. In \cite{erignouxHydrodynamicLimitActive2021} it was not shown that $\pi_t( \dd \x , \dd \theta ) = f_t( \x ,\theta) \dd \x \dd\theta$, but rather that there exists for any $t,\;x$ an orientation distribution $\hat{f}_t(\x,\dd\theta)$ on $\SS$ such that $\pi_t( \dd \x , \dd \theta )=\hat{f}_t(\x, \dd\theta)\dd \x$. Here we chose, to make it more readable, to write statement $i)$ of Proposition \ref{prop:relcomp} this way, assuming that $\pi_t$ is, at any $\x$, absolutely continuous with respect to the Lebesgue measure on $\SS$.

However, this is not a problem because our orientation dynamics are given by angular diffusion, and the absolute continuity of $\pi_t$ on $\SS$ can be straightforwardly shown in two ways.
\begin{enumerate}
\item Either by showing directly on the microscopic system that, since particle's angles diffuse independently, the distribution of angles $\hat \rho_{\x}^{\varepsilon N}$ in a mesoscopic box of size $\ep N$ must be, for any positive time, absolutely continuous w.r.t. the Lebesgue measure for any $t$, $\x$.
\item Or by proving the statement at a macroscopic level, by first showing a weakened hydrodynamic limit taking the form $\hat{f}_t(\x, \dd\theta)$ in the spirit of \cite{erignouxHydrodynamicLimitActive2021}, and then using the maximum principle on the orientation diffusion to show that any such weak solution $\hat{f}_t(\x, \dd\theta)$ is actually continuous with respect to the Lebesgue measure and can therefore be written $\hat{f}_t(\x, \dd\theta)=f_t(\x, \theta)\dd\theta$
\end{enumerate}
Note that this was not true for the {\macc} studied in \cite{erignouxHydrodynamicLimitActive2021} because the dynamics considered there was a jump dynamics with alignment, which had no such instantaneous mixing properties on particles' orientations. The fact that  $\hat{f}_t(\x,\dd\theta)$ is absolutely continuous with respect to the Lebesgue measure on $\SS$ is, therefore, a direct consequence of the angular diffusion dynamics we chose for particle's orientations in our {\macc}. 

\end{remark}

\section{Discussion}
In this paper, we considered a lattice model for active matter and derived its exact hydrodynamic limit. This model exhibits a phenomenon of motility-induced phase separation (MIPS), which results from the interplay between self-propulsion and particle crowding. When self-propelled particles form clusters, they tend to slow down due to collisions and interactions. This creates a positive feedback loop that enhances the clustering and leads to a phase separation between a dense (liquid-like) phase and a dilute (gas-like) phase. 

We obtain an exact macroscopic description of the active lattice gas model in terms of a PDE and use it to characterise the large-scale behaviours that emerge from the microscopic dynamics. Note that the hydrodynamic limit derived in Section \ref{sec_hydrolim} is exact for any finite time in the limit $N \to \infty$. However, as the numerical simulations show, large but finite-sized systems are subject to fluctuations around the hydrodynamic limit. Macroscopic fluctuation theory (or Large deviation theory) is required to account for such effects \cite{quastelLargeDeviationsSymmetric1999,agranovEntropyProductionIts2022,erignouxHydrodynamicsActiveMatter2021}.  These terms are needed to understand metastable solutions, but their rigorous derivation remains an open problem. 

Our analysis of MIPS consists of two approaches: linear stability analysis of the PDE model and numerical simulations with perturbed initial conditions. %Linear stability analysis was used to find an upper bound in phase space for which the stationary state is stable. 
The latter approach allows us to observe the emergence and evolution of different patterns from the homogeneous state. Unlike previous studies \cite{catesMotilityInducedPhaseSeparation2015,kourbane-housseneExactHydrodynamicDescription2018} that obtained analytical solutions for the stationary states, we resort to numerical methods due to the complexity of our model, which involves nonlinear coefficients and arbitrary orientations. 

While most previous work on hydrodynamic limits focuses on gradient systems \cite{kourbane-housseneExactHydrodynamicDescription2018,goncalvesHydrodynamicLimitsEmergence2023}, we do expect non-gradient systems to be ubiquitous, for example, in multi-component systems with simple exclusion \cite{berendsenCrossdiffusionModelMultiple2017, masonMacroscopicBehaviourTwoSpecies2023} but also elsewhere, such as in systems with general exclusion \cite{aritaVariationalCalculationTransport2017}. Among such systems, the one we have here is well-chosen because we can handle non-explicit coefficients stemming from the non-gradient aspect via our knowledge of the self-diffusion coefficient $d_s(\rho)$.

Our results demonstrate the complexity and richness of active matter systems and highlight several promising avenues for future research. 
For example, the phenomenology of MIPS usually includes metastable behaviour between the spinodal and binodal~\cite{catesMotilityInducedPhaseSeparation2015}, in which case \eqref{act_pas_eq} will support both homogeneous and inhomogeneous steady solutions and transitions between these states would require an analysis of large deviations. Phase separation occurs in the limit of fast rotational dynamics that cause the interfacial widths to shrink to zero. This limit leads to a macroscopic description in terms of the density $\rho$ alone (instead of the orientation-dependence density $f$): it would be useful to have a rigorous mathematical formulation of this limit. %Finally, we recall the deviations in Fig.~\ref{fig_1} between the limit of stability and the result of the linear theory: it would be interesting to characterise the mathematics of this nonlinear instability in more detail and to understand its physical origin.
 
\bibliography{refs.bib}


\begin{thebibliography}{58}
\ifx \bisbn   \undefined \def \bisbn  #1{ISBN #1}\fi
\ifx \binits  \undefined \def \binits#1{#1}\fi
\ifx \bauthor  \undefined \def \bauthor#1{#1}\fi
\ifx \batitle  \undefined \def \batitle#1{#1}\fi
\ifx \bjtitle  \undefined \def \bjtitle#1{#1}\fi
\ifx \bvolume  \undefined \def \bvolume#1{\textbf{#1}}\fi
\ifx \byear  \undefined \def \byear#1{#1}\fi
\ifx \bissue  \undefined \def \bissue#1{#1}\fi
\ifx \bfpage  \undefined \def \bfpage#1{#1}\fi
\ifx \blpage  \undefined \def \blpage #1{#1}\fi
\ifx \burl  \undefined \def \burl#1{\textsf{#1}}\fi
\ifx \doiurl  \undefined \def \doiurl#1{\url{https://doi.org/#1}}\fi
\ifx \betal  \undefined \def \betal{\textit{et al.}}\fi
\ifx \binstitute  \undefined \def \binstitute#1{#1}\fi
\ifx \binstitutionaled  \undefined \def \binstitutionaled#1{#1}\fi
\ifx \bctitle  \undefined \def \bctitle#1{#1}\fi
\ifx \beditor  \undefined \def \beditor#1{#1}\fi
\ifx \bpublisher  \undefined \def \bpublisher#1{#1}\fi
\ifx \bbtitle  \undefined \def \bbtitle#1{#1}\fi
\ifx \bedition  \undefined \def \bedition#1{#1}\fi
\ifx \bseriesno  \undefined \def \bseriesno#1{#1}\fi
\ifx \blocation  \undefined \def \blocation#1{#1}\fi
\ifx \bsertitle  \undefined \def \bsertitle#1{#1}\fi
\ifx \bsnm \undefined \def \bsnm#1{#1}\fi
\ifx \bsuffix \undefined \def \bsuffix#1{#1}\fi
\ifx \bparticle \undefined \def \bparticle#1{#1}\fi
\ifx \barticle \undefined \def \barticle#1{#1}\fi
\bibcommenthead
\ifx \bconfdate \undefined \def \bconfdate #1{#1}\fi
\ifx \botherref \undefined \def \botherref #1{#1}\fi
\ifx \url \undefined \def \url#1{\textsf{#1}}\fi
\ifx \bchapter \undefined \def \bchapter#1{#1}\fi
\ifx \bbook \undefined \def \bbook#1{#1}\fi
\ifx \bcomment \undefined \def \bcomment#1{#1}\fi
\ifx \oauthor \undefined \def \oauthor#1{#1}\fi
\ifx \citeauthoryear \undefined \def \citeauthoryear#1{#1}\fi
\ifx \endbibitem  \undefined \def \endbibitem {}\fi
\ifx \bconflocation  \undefined \def \bconflocation#1{#1}\fi
\ifx \arxivurl  \undefined \def \arxivurl#1{\textsf{#1}}\fi
\csname PreBibitemsHook\endcsname

%%% 1
\bibitem[\protect\citeauthoryear{Vicsek and
  Zafeiris}{2012}]{vicsekCollectiveMotion2012}
\begin{barticle}
\bauthor{\bsnm{Vicsek}, \binits{T.}},
\bauthor{\bsnm{Zafeiris}, \binits{A.}}:
\batitle{Collective motion}.
\bjtitle{Physics Reports}
\bvolume{517}(\bissue{3}),
\bfpage{71}--\blpage{140}
(\byear{2012})
\doiurl{10.1016/j.physrep.2012.03.004}
\end{barticle}
\endbibitem

%%% 2
\bibitem[\protect\citeauthoryear{Burger
  et~al.}{2016}]{burgerLaneFormationSideStepping2016}
\begin{barticle}
\bauthor{\bsnm{Burger}, \binits{M.}},
\bauthor{\bsnm{Hittmeir}, \binits{S.}},
\bauthor{\bsnm{Ranetbauer}, \binits{H.}},
\bauthor{\bsnm{Wolfram}, \binits{M.-T.}}:
\batitle{Lane {{Formation}} by {{Side-Stepping}}}.
\bjtitle{SIAM Journal on Mathematical Analysis}
\bvolume{48}(\bissue{2}),
\bfpage{981}--\blpage{1005}
(\byear{2016})
\doiurl{10.1137/15M1033174}
\end{barticle}
\endbibitem

%%% 3
\bibitem[\protect\citeauthoryear{Bacik
  et~al.}{2023}]{bacikLaneNucleationComplex2023}
\begin{barticle}
\bauthor{\bsnm{Bacik}, \binits{K.A.}},
\bauthor{\bsnm{Bacik}, \binits{B.S.}},
\bauthor{\bsnm{Rogers}, \binits{T.}}:
\batitle{Lane nucleation in complex active flows}.
\bjtitle{Science}
\bvolume{379}(\bissue{6635}),
\bfpage{923}--\blpage{928}
(\byear{2023})
\doiurl{10.1126/science.add8091}
\end{barticle}
\endbibitem

%%% 4
\bibitem[\protect\citeauthoryear{Buttinoni
  et~al.}{2013}]{buttinoniDynamicalClusteringPhase2013}
\begin{barticle}
\bauthor{\bsnm{Buttinoni}, \binits{I.}},
\bauthor{\bsnm{Bialk{\'e}}, \binits{J.}},
\bauthor{\bsnm{K{\"u}mmel}, \binits{F.}},
\bauthor{\bsnm{L{\"o}wen}, \binits{H.}},
\bauthor{\bsnm{Bechinger}, \binits{C.}},
\bauthor{\bsnm{Speck}, \binits{T.}}:
\batitle{Dynamical {{Clustering}} and {{Phase Separation}} in {{Suspensions}}
  of {{Self-Propelled Colloidal Particles}}}.
\bjtitle{Physical Review Letters}
\bvolume{110}(\bissue{23}),
\bfpage{238301}
(\byear{2013})
\doiurl{10.1103/PhysRevLett.110.238301}
\end{barticle}
\endbibitem

%%% 5
\bibitem[\protect\citeauthoryear{Fily and
  Marchetti}{2012}]{filyAthermalPhaseSeparation2012}
\begin{barticle}
\bauthor{\bsnm{Fily}, \binits{Y.}},
\bauthor{\bsnm{Marchetti}, \binits{M.C.}}:
\batitle{Athermal {{Phase Separation}} of {{Self-Propelled Particles}} with
  {{No Alignment}}}.
\bjtitle{Physical Review Letters}
\bvolume{108}(\bissue{23}),
\bfpage{235702}
(\byear{2012})
\doiurl{10.1103/PhysRevLett.108.235702}
\end{barticle}
\endbibitem

%%% 6
\bibitem[\protect\citeauthoryear{Redner
  et~al.}{2013}]{rednerStructureDynamicsPhaseSeparating2013}
\begin{barticle}
\bauthor{\bsnm{Redner}, \binits{G.}},
\bauthor{\bsnm{Hagan}, \binits{M.}},
\bauthor{\bsnm{Baskaran}, \binits{A.}}:
\batitle{Structure and {{Dynamics}} of a {{Phase-Separating Active Colloidal
  Fluid}}}.
\bjtitle{Physical review letters}
\bvolume{110},
\bfpage{055701}
(\byear{2013})
\doiurl{10.1103/PhysRevLett.110.055701}
\end{barticle}
\endbibitem

%%% 7
\bibitem[\protect\citeauthoryear{B{\"a}r
  et~al.}{2020}]{barSelfPropelledRodsInsights2020}
\begin{barticle}
\bauthor{\bsnm{B{\"a}r}, \binits{M.}},
\bauthor{\bsnm{Gro{\ss}mann}, \binits{R.}},
\bauthor{\bsnm{Heidenreich}, \binits{S.}},
\bauthor{\bsnm{Peruani}, \binits{F.}}:
\batitle{Self-{{Propelled Rods}}: {{Insights}} and {{Perspectives}} for
  {{Active Matter}}}.
\bjtitle{Annual Review of Condensed Matter Physics}
\bvolume{11}(\bissue{1}),
\bfpage{441}--\blpage{466}
(\byear{2020})
\doiurl{10.1146/annurev-conmatphys-031119-050611}
\end{barticle}
\endbibitem

%%% 8
\bibitem[\protect\citeauthoryear{Sokolov and
  Aranson}{2012}]{sokolovPhysicalPropertiesCollective2012}
\begin{barticle}
\bauthor{\bsnm{Sokolov}, \binits{A.}},
\bauthor{\bsnm{Aranson}, \binits{I.S.}}:
\batitle{Physical {{Properties}} of {{Collective Motion}} in {{Suspensions}} of
  {{Bacteria}}}.
\bjtitle{Physical Review Letters}
\bvolume{109}(\bissue{24}),
\bfpage{248109}
(\byear{2012})
\doiurl{10.1103/PhysRevLett.109.248109}
\end{barticle}
\endbibitem

%%% 9
\bibitem[\protect\citeauthoryear{Berg}{1993}]{bergRandomWalksBiology1993}
\begin{bbook}
\bauthor{\bsnm{Berg}, \binits{H.C.}}:
\bbtitle{Random {{Walks}} in {{Biology}}}.
\bpublisher{{Princeton University Press}},
\blocation{{Princeton, NJ}}
(\byear{1993})
\end{bbook}
\endbibitem

%%% 10
\bibitem[\protect\citeauthoryear{Sumino
  et~al.}{2012}]{suminoLargescaleVortexLattice2012}
\begin{barticle}
\bauthor{\bsnm{Sumino}, \binits{Y.}},
\bauthor{\bsnm{Nagai}, \binits{K.H.}},
\bauthor{\bsnm{Shitaka}, \binits{Y.}},
\bauthor{\bsnm{Tanaka}, \binits{D.}},
\bauthor{\bsnm{Yoshikawa}, \binits{K.}},
\bauthor{\bsnm{Chat{\'e}}, \binits{H.}},
\bauthor{\bsnm{Oiwa}, \binits{K.}}:
\batitle{Large-scale vortex lattice emerging from collectively moving
  microtubules}.
\bjtitle{Nature}
\bvolume{483}(\bissue{7390}),
\bfpage{448}--\blpage{452}
(\byear{2012})
\doiurl{10.1038/nature10874}
\end{barticle}
\endbibitem

%%% 11
\bibitem[\protect\citeauthoryear{Giardina}{2008}]{giardinaCollectiveBehaviorAnimal2008}
\begin{barticle}
\bauthor{\bsnm{Giardina}, \binits{I.}}:
\batitle{Collective behavior in animal groups: {{Theoretical}} models and
  empirical studies}.
\bjtitle{HFSP Journal}
\bvolume{2}(\bissue{4}),
\bfpage{205}--\blpage{219}
(\byear{2008})
\doiurl{10.2976/1.2961038}
\end{barticle}
\endbibitem

%%% 12
\bibitem[\protect\citeauthoryear{Cates and
  Tailleur}{2015}]{catesMotilityInducedPhaseSeparation2015}
\begin{barticle}
\bauthor{\bsnm{Cates}, \binits{M.E.}},
\bauthor{\bsnm{Tailleur}, \binits{J.}}:
\batitle{Motility-{{Induced Phase Separation}}}.
\bjtitle{Annual Review of Condensed Matter Physics}
\bvolume{6}(\bissue{1}),
\bfpage{219}--\blpage{244}
(\byear{2015})
\doiurl{10.1146/annurev-conmatphys-031214-014710}
\end{barticle}
\endbibitem

%%% 13
\bibitem[\protect\citeauthoryear{Stenhammar
  et~al.}{2015}]{stenhammarActivityInducedPhaseSeparation2015}
\begin{barticle}
\bauthor{\bsnm{Stenhammar}, \binits{J.}},
\bauthor{\bsnm{Wittkowski}, \binits{R.}},
\bauthor{\bsnm{Marenduzzo}, \binits{D.}},
\bauthor{\bsnm{Cates}, \binits{M.E.}}:
\batitle{Activity-{{Induced Phase Separation}} and {{Self-Assembly}} in
  {{Mixtures}} of {{Active}} and {{Passive Particles}}}.
\bjtitle{Physical Review Letters}
\bvolume{114}(\bissue{1}),
\bfpage{018301}
(\byear{2015})
\doiurl{10.1103/PhysRevLett.114.018301}
\end{barticle}
\endbibitem

%%% 14
\bibitem[\protect\citeauthoryear{Solon
  et~al.}{2018}]{solonGeneralizedThermodynamicsMotilityinduced2018}
\begin{barticle}
\bauthor{\bsnm{Solon}, \binits{A.P.}},
\bauthor{\bsnm{Stenhammar}, \binits{J.}},
\bauthor{\bsnm{Cates}, \binits{M.E.}},
\bauthor{\bsnm{Kafri}, \binits{Y.}},
\bauthor{\bsnm{Tailleur}, \binits{J.}}:
\batitle{Generalized thermodynamics of motility-induced phase separation: Phase
  equilibria, {{Laplace}} pressure, and change of ensembles}.
\bjtitle{New Journal of Physics}
\bvolume{20}(\bissue{7}),
\bfpage{075001}
(\byear{2018})
\doiurl{10.1088/1367-2630/aaccdd}
\end{barticle}
\endbibitem

%%% 15
\bibitem[\protect\citeauthoryear{Wittkowski
  et~al.}{2017}]{wittkowskiNonequilibriumDynamicsMixtures2017}
\begin{barticle}
\bauthor{\bsnm{Wittkowski}, \binits{R.}},
\bauthor{\bsnm{Stenhammar}, \binits{J.}},
\bauthor{\bsnm{Cates}, \binits{M.E.}}:
\batitle{Nonequilibrium dynamics of mixtures of active and passive colloidal
  particles}.
\bjtitle{New Journal of Physics}
\bvolume{19}(\bissue{10}),
\bfpage{105003}
(\byear{2017})
\doiurl{10.1088/1367-2630/aa8195}
\end{barticle}
\endbibitem

%%% 16
\bibitem[\protect\citeauthoryear{Bruna
  et~al.}{2022}]{brunaPhaseSeparationSystems2022}
\begin{barticle}
\bauthor{\bsnm{Bruna}, \binits{M.}},
\bauthor{\bsnm{Burger}, \binits{M.}},
\bauthor{\bsnm{Esposito}, \binits{A.}},
\bauthor{\bsnm{Schulz}, \binits{S.M.}}:
\batitle{Phase {{Separation}} in {{Systems}} of {{Interacting Active Brownian
  Particles}}}.
\bjtitle{SIAM Journal on Applied Mathematics}
\bvolume{82}(\bissue{4}),
\bfpage{1635}--\blpage{1660}
(\byear{2022})
\doiurl{10.1137/21M1452524}
\end{barticle}
\endbibitem

%%% 17
\bibitem[\protect\citeauthoryear{Wysocki
  et~al.}{2016}]{wysockiPropagatingInterfacesMixtures2016}
\begin{barticle}
\bauthor{\bsnm{Wysocki}, \binits{A.}},
\bauthor{\bsnm{Winkler}, \binits{R.G.}},
\bauthor{\bsnm{Gompper}, \binits{G.}}:
\batitle{Propagating interfaces in mixtures of active and passive {{Brownian}}
  particles}.
\bjtitle{New Journal of Physics}
\bvolume{18}(\bissue{12}),
\bfpage{123030}
(\byear{2016})
\doiurl{10.1088/1367-2630/aa529d}
\end{barticle}
\endbibitem

%%% 18
\bibitem[\protect\citeauthoryear{Bialk{\'e}
  et~al.}{2013}]{bialkeMicroscopicTheoryPhase2013}
\begin{barticle}
\bauthor{\bsnm{Bialk{\'e}}, \binits{J.}},
\bauthor{\bsnm{L{\"o}wen}, \binits{H.}},
\bauthor{\bsnm{Speck}, \binits{T.}}:
\batitle{Microscopic theory for the phase separation of self-propelled
  repulsive disks}.
\bjtitle{EPL (Europhysics Letters)}
\bvolume{103},
\bfpage{30008}
(\byear{2013})
\doiurl{10.1209/0295-5075/103/30008}
\end{barticle}
\endbibitem

%%% 19
\bibitem[\protect\citeauthoryear{Speck
  et~al.}{2015}]{speckDynamicalMeanfieldTheory2015}
\begin{barticle}
\bauthor{\bsnm{Speck}, \binits{T.}},
\bauthor{\bsnm{Menzel}, \binits{A.M.}},
\bauthor{\bsnm{Bialk{\'e}}, \binits{J.}},
\bauthor{\bsnm{L{\"o}wen}, \binits{H.}}:
\batitle{Dynamical mean-field theory and weakly non-linear analysis for the
  phase separation of active {{Brownian}} particles}.
\bjtitle{The Journal of Chemical Physics}
\bvolume{142}(\bissue{22}),
\bfpage{224109}
(\byear{2015})
\doiurl{10.1063/1.4922324}
\end{barticle}
\endbibitem

%%% 20
\bibitem[\protect\citeauthoryear{Tailleur and
  Cates}{2008}]{tailleurStatisticalMechanicsInteracting2008}
\begin{barticle}
\bauthor{\bsnm{Tailleur}, \binits{J.}},
\bauthor{\bsnm{Cates}, \binits{M.E.}}:
\batitle{Statistical {{Mechanics}} of {{Interacting Run-and-Tumble Bacteria}}}.
\bjtitle{Physical Review Letters}
\bvolume{100}(\bissue{21}),
\bfpage{218103}
(\byear{2008})
\doiurl{10.1103/PhysRevLett.100.218103}
\end{barticle}
\endbibitem

%%% 21
\bibitem[\protect\citeauthoryear{Wittkowski
  et~al.}{2014}]{wittkowskiScalarF4Field2014}
\begin{barticle}
\bauthor{\bsnm{Wittkowski}, \binits{R.}},
\bauthor{\bsnm{Tiribocchi}, \binits{A.}},
\bauthor{\bsnm{Stenhammar}, \binits{J.}},
\bauthor{\bsnm{Allen}, \binits{R.J.}},
\bauthor{\bsnm{Marenduzzo}, \binits{D.}},
\bauthor{\bsnm{Cates}, \binits{M.E.}}:
\batitle{Scalar {$\Phi$}4 field theory for active-particle phase separation}.
\bjtitle{Nature Communications}
\bvolume{5}(\bissue{1}),
\bfpage{4351}
(\byear{2014})
\doiurl{10.1038/ncomms5351}
\end{barticle}
\endbibitem

%%% 22
\bibitem[\protect\citeauthoryear{Bruna
  et~al.}{2023}]{brunaDerivationMacroscopicModel2023}
\begin{barticle}
\bauthor{\bsnm{Bruna}, \binits{M.}},
\bauthor{\bsnm{Chapman}, \binits{S.J.}},
\bauthor{\bsnm{Schmidtchen}, \binits{M.}}:
\batitle{Derivation of a macroscopic model for {{Brownian}} hard needles}.
\bjtitle{Proceedings of the Royal Society A: Mathematical, Physical and
  Engineering Sciences}
\bvolume{479}(\bissue{2274}),
\bfpage{20230076}
(\byear{2023})
\doiurl{10.1098/rspa.2023.0076}
\end{barticle}
\endbibitem

%%% 23
\bibitem[\protect\citeauthoryear{{Kourbane-Houssene}
  et~al.}{2018}]{kourbane-housseneExactHydrodynamicDescription2018}
\begin{barticle}
\bauthor{\bsnm{{Kourbane-Houssene}}, \binits{M.}},
\bauthor{\bsnm{Erignoux}, \binits{C.}},
\bauthor{\bsnm{Bodineau}, \binits{T.}},
\bauthor{\bsnm{Tailleur}, \binits{J.}}:
\batitle{Exact {{Hydrodynamic Description}} of {{Active Lattice Gases}}}.
\bjtitle{Physical Review Letters}
\bvolume{120}(\bissue{26}),
\bfpage{268003}
(\byear{2018})
\doiurl{10.1103/PhysRevLett.120.268003}
\end{barticle}
\endbibitem

%%% 24
\bibitem[\protect\citeauthoryear{Erignoux}{2021}]{erignouxHydrodynamicsActiveMatter2021}
\begin{botherref}
\oauthor{\bsnm{Erignoux}, \binits{C.}}:
On the Hydrodynamics of Active Matter Models on a Lattice.
{arXiv}
(2021).
\doiurl{10.48550/arXiv.2108.04003}
\end{botherref}
\endbibitem

%%% 25
\bibitem[\protect\citeauthoryear{Kipnis and
  Landim}{1998}]{kipnisScalingLimitsInteracting1998}
\begin{bbook}
\bauthor{\bsnm{Kipnis}, \binits{C.}},
\bauthor{\bsnm{Landim}, \binits{C.}}:
\bbtitle{Scaling {{Limits}} of {{Interacting Particle Systems}}}.
\bpublisher{{Springer Science \& Business Media}},
\blocation{{Berlin, Heidelberg}}
(\byear{1998})
\end{bbook}
\endbibitem

%%% 26
\bibitem[\protect\citeauthoryear{Quastel}{1992}]{quastelDiffusionColorSimple1992}
\begin{barticle}
\bauthor{\bsnm{Quastel}, \binits{J.}}:
\batitle{Diffusion of color in the simple exclusion process}.
\bjtitle{Communications on Pure and Applied Mathematics}
\bvolume{45}(\bissue{6}),
\bfpage{623}--\blpage{679}
(\byear{1992})
\doiurl{10.1002/cpa.3160450602}
\end{barticle}
\endbibitem

%%% 27
\bibitem[\protect\citeauthoryear{Erignoux}{2021}]{erignouxHydrodynamicLimitActive2021}
\begin{botherref}
\oauthor{\bsnm{Erignoux}, \binits{C.}}:
Hydrodynamic limit for an active exclusion process.
M\'emoires de la Soci\'et\'e Math\'ematique de France
\textbf{169}
(2021)
\doiurl{10.24033/msmf.477}
\end{botherref}
\endbibitem

%%% 28
\bibitem[\protect\citeauthoryear{Guo
  et~al.}{1988}]{guoNonlinearDiffusionLimit1988}
\begin{barticle}
\bauthor{\bsnm{Guo}, \binits{M.Z.}},
\bauthor{\bsnm{Papanicolaou}, \binits{G.C.}},
\bauthor{\bsnm{Varadhan}, \binits{S.R.S.}}:
\batitle{Nonlinear diffusion limit for a system with nearest neighbor
  interactions}.
\bjtitle{Communications in Mathematical Physics}
\bvolume{118}(\bissue{1}),
\bfpage{31}--\blpage{59}
(\byear{1988})
\end{barticle}
\endbibitem

%%% 29
\bibitem[\protect\citeauthoryear{Spohn}{1990}]{spohnTracerDiffusionLattice1990}
\begin{barticle}
\bauthor{\bsnm{Spohn}, \binits{H.}}:
\batitle{Tracer diffusion in lattice gases}.
\bjtitle{Journal of Statistical Physics}
\bvolume{59}(\bissue{5}),
\bfpage{1227}--\blpage{1239}
(\byear{1990})
\doiurl{10.1007/BF01334748}
\end{barticle}
\endbibitem

%%% 30
\bibitem[\protect\citeauthoryear{Varadhan}{1994}]{varadhanRegularitySelfDiffusionCoefficient1994}
\begin{bchapter}
\bauthor{\bsnm{Varadhan}, \binits{S.R.S.}}:
\bctitle{Regularity of {{Self-Diffusion Coefficient}}}.
In: \beditor{\bsnm{Freidlin}, \binits{M.I.}} (ed.)
\bbtitle{The {{Dynkin Festschrift}}}.
\bsertitle{Progress in {{Probability}}},
vol. \bseriesno{34},
pp. \bfpage{387}--\blpage{397}.
\bpublisher{{Birkh\"auser}},
\blocation{{Boston, MA}}
(\byear{1994})
\end{bchapter}
\endbibitem

%%% 31
\bibitem[\protect\citeauthoryear{Landim
  et~al.}{2001}]{landimSymmetricSimpleExclusion2001}
\begin{barticle}
\bauthor{\bsnm{Landim}, \binits{C.}},
\bauthor{\bsnm{Olla}, \binits{S.}},
\bauthor{\bsnm{Varadhan}, \binits{S.R.S.}}:
\batitle{Symmetric {{Simple Exclusion Process}}:\textparagraph{{Regularity}} of
  the {{Self-Diffusion Coefficient}}}.
\bjtitle{Communications in Mathematical Physics}
\bvolume{224}(\bissue{1}),
\bfpage{307}--\blpage{321}
(\byear{2001})
\doiurl{10.1007/s002200100513}
\end{barticle}
\endbibitem

%%% 32
\bibitem[\protect\citeauthoryear{Barbaro
  et~al.}{2016}]{barbaroPhaseTransitionsKinetic2016}
\begin{barticle}
\bauthor{\bsnm{Barbaro}, \binits{A.B.T.}},
\bauthor{\bsnm{Cannizo}, \binits{J.A.}},
\bauthor{\bsnm{Carrillo}, \binits{J.A.}},
\bauthor{\bsnm{Degond}, \binits{P.}}:
\batitle{Phase {{Transitions}} in a {{Kinetic Flocking Model}} of
  {{Cucker--Smale Type}}}.
\bjtitle{Multiscale Modeling \& Simulation}
\bvolume{14}(\bissue{3}),
\bfpage{1063}--\blpage{1088}
(\byear{2016})
\doiurl{10.1137/15M1043637}
\end{barticle}
\endbibitem

%%% 33
\bibitem[\protect\citeauthoryear{Mielke
  et~al.}{2014}]{mielkeRelationGradientFlows2014}
\begin{barticle}
\bauthor{\bsnm{Mielke}, \binits{A.}},
\bauthor{\bsnm{Peletier}, \binits{M.A.}},
\bauthor{\bsnm{Renger}, \binits{D.R.M.}}:
\batitle{On the {{Relation}} between {{Gradient Flows}} and the
  {{Large-Deviation Principle}}, with {{Applications}} to {{Markov Chains}} and
  {{Diffusion}}}.
\bjtitle{Potential Analysis}
\bvolume{41}(\bissue{4}),
\bfpage{1293}--\blpage{1327}
(\byear{2014})
\doiurl{10.1007/s11118-014-9418-5}
\end{barticle}
\endbibitem

%%% 34
\bibitem[\protect\citeauthoryear{Mason
  et~al.}{2023}]{masonMacroscopicBehaviourTwoSpecies2023}
\begin{barticle}
\bauthor{\bsnm{Mason}, \binits{J.}},
\bauthor{\bsnm{Jack}, \binits{R.L.}},
\bauthor{\bsnm{Bruna}, \binits{M.}}:
\batitle{Macroscopic {{Behaviour}} in a {{Two-Species Exclusion Process Via}}
  the {{Method}} of {{Matched Asymptotics}}}.
\bjtitle{Journal of Statistical Physics}
\bvolume{190}(\bissue{3}),
\bfpage{47}
(\byear{2023})
\doiurl{10.1007/s10955-022-03036-9}
\end{barticle}
\endbibitem

%%% 35
\bibitem[\protect\citeauthoryear{Nakazato and
  Kitahara}{1980}]{nakazatoSiteBlockingEffect1980}
\begin{barticle}
\bauthor{\bsnm{Nakazato}, \binits{K.}},
\bauthor{\bsnm{Kitahara}, \binits{K.}}:
\batitle{Site {{Blocking Effect}} in {{Tracer Diffusion}} on a {{Lattice}}}.
\bjtitle{Progress of Theoretical Physics}
\bvolume{64}(\bissue{6}),
\bfpage{2261}--\blpage{2264}
(\byear{1980})
\doiurl{10.1143/PTP.64.2261}
\end{barticle}
\endbibitem

%%% 36
\bibitem[\protect\citeauthoryear{Illien
  et~al.}{2018}]{illienNonequilibriumFluctuationsEnhanced2018}
\begin{barticle}
\bauthor{\bsnm{Illien}, \binits{P.}},
\bauthor{\bsnm{B{\'e}nichou}, \binits{O.}},
\bauthor{\bsnm{Oshanin}, \binits{G.}},
\bauthor{\bsnm{Sarracino}, \binits{A.}},
\bauthor{\bsnm{Voituriez}, \binits{R.}}:
\batitle{Nonequilibrium {{Fluctuations}} and {{Enhanced Diffusion}} of a
  {{Driven Particle}} in a {{Dense Environment}}}.
\bjtitle{Physical Review Letters}
\bvolume{120}(\bissue{20}),
\bfpage{200606}
(\byear{2018})
\doiurl{10.1103/PhysRevLett.120.200606}
\end{barticle}
\endbibitem

%%% 37
\bibitem[\protect\citeauthoryear{Rizkallah
  et~al.}{2022}]{rizkallahMicroscopicTheoryDiffusion2022}
\begin{barticle}
\bauthor{\bsnm{Rizkallah}, \binits{P.}},
\bauthor{\bsnm{Sarracino}, \binits{A.}},
\bauthor{\bsnm{B{\'e}nichou}, \binits{O.}},
\bauthor{\bsnm{Illien}, \binits{P.}}:
\batitle{Microscopic {{Theory}} for the {{Diffusion}} of an {{Active Particle}}
  in a {{Crowded Environment}}}.
\bjtitle{Physical Review Letters}
\bvolume{128}(\bissue{3}),
\bfpage{038001}
(\byear{2022})
\doiurl{10.1103/PhysRevLett.128.038001}
\end{barticle}
\endbibitem

%%% 38
\bibitem[\protect\citeauthoryear{Arita
  et~al.}{2014}]{aritaGeneralizedExclusionProcesses2014}
\begin{barticle}
\bauthor{\bsnm{Arita}, \binits{C.}},
\bauthor{\bsnm{Krapivsky}, \binits{P.L.}},
\bauthor{\bsnm{Mallick}, \binits{K.}}:
\batitle{Generalized exclusion processes: {{Transport}} coefficients}.
\bjtitle{Physical Review E}
\bvolume{90}(\bissue{5}),
\bfpage{052108}
(\byear{2014})
\doiurl{10.1103/PhysRevE.90.052108}
\end{barticle}
\endbibitem

%%% 39
\bibitem[\protect\citeauthoryear{Arita
  et~al.}{2018}]{aritaBulkDiffusionKinetically2018}
\begin{barticle}
\bauthor{\bsnm{Arita}, \binits{C.}},
\bauthor{\bsnm{Krapivsky}, \binits{P.L.}},
\bauthor{\bsnm{Mallick}, \binits{K.}}:
\batitle{Bulk diffusion in a kinetically constrained lattice gas}.
\bjtitle{Journal of Physics A: Mathematical and Theoretical}
\bvolume{51}(\bissue{12}),
\bfpage{125002}
(\byear{2018})
\doiurl{10.1088/1751-8121/aaac89}
\end{barticle}
\endbibitem

%%% 40
\bibitem[\protect\citeauthoryear{Dabaghi
  et~al.}{2022}]{dabaghiComputationSelfdiffusionCoefficient2022}
\begin{botherref}
\oauthor{\bsnm{Dabaghi}, \binits{J.}},
\oauthor{\bsnm{Ehrlacher}, \binits{V.}},
\oauthor{\bsnm{Str{\"o}ssner}, \binits{C.}}:
Computation of the Self-Diffusion Coefficient with Low-Rank Tensor Methods:
  Application to the Simulation of a Cross-Diffusion System.
{arXiv}
(2022).
\doiurl{10.48550/arXiv.2111.11349}
\end{botherref}
\endbibitem

%%% 41
\bibitem[\protect\citeauthoryear{Dabaghi
  et~al.}{2023}]{dabaghiTensorApproximationSelfdiffusion2023}
\begin{barticle}
\bauthor{\bsnm{Dabaghi}, \binits{J.}},
\bauthor{\bsnm{Ehrlacher}, \binits{V.}},
\bauthor{\bsnm{Str{\"o}ssner}, \binits{C.}}:
\batitle{Tensor approximation of the self-diffusion matrix of tagged particle
  processes}.
\bjtitle{Journal of Computational Physics}
\bvolume{480},
\bfpage{112017}
(\byear{2023})
\doiurl{10.1016/j.jcp.2023.112017}
\end{barticle}
\endbibitem

%%% 42
\bibitem[\protect\citeauthoryear{Ziener
  et~al.}{2012}]{zienerMathieuFunctionsPurely2012}
\begin{barticle}
\bauthor{\bsnm{Ziener}, \binits{C.H.}},
\bauthor{\bsnm{R{\"u}ckl}, \binits{M.}},
\bauthor{\bsnm{Kampf}, \binits{T.}},
\bauthor{\bsnm{Bauer}, \binits{W.R.}},
\bauthor{\bsnm{Schlemmer}, \binits{H.P.}}:
\batitle{Mathieu functions for purely imaginary parameters}.
\bjtitle{Journal of Computational and Applied Mathematics}
\bvolume{236}(\bissue{17}),
\bfpage{4513}--\blpage{4524}
(\byear{2012})
\doiurl{10.1016/j.cam.2012.04.023}
\end{barticle}
\endbibitem

%%% 43
\bibitem[\protect\citeauthoryear{Gautschi}{1967}]{gautschiComputationalAspectsThreeTerm1967}
\begin{barticle}
\bauthor{\bsnm{Gautschi}, \binits{W.}}:
\batitle{Computational {{Aspects}} of {{Three-Term Recurrence Relations}}}.
\bjtitle{SIAM Review}
\bvolume{9}(\bissue{1}),
\bfpage{24}--\blpage{82}
(\byear{1967})
\doiurl{10.1137/1009002}
\end{barticle}
\endbibitem

%%% 44
\bibitem[\protect\citeauthoryear{Ikebe
  et~al.}{1996}]{ikebeEigenvalueProblemInfinite1996}
\begin{barticle}
\bauthor{\bsnm{Ikebe}, \binits{Y.}},
\bauthor{\bsnm{Asai}, \binits{N.}},
\bauthor{\bsnm{Miyazaki}, \binits{Y.}},
\bauthor{\bsnm{Cai}, \binits{D.}}:
\batitle{The eigenvalue problem for infinite complex symmetric tridiagonal
  matrices with application}.
\bjtitle{Linear Algebra and its Applications}
\bvolume{241--243},
\bfpage{599}--\blpage{618}
(\byear{1996})
\doiurl{10.1016/0024-3795(95)00699-0}
\end{barticle}
\endbibitem

%%% 45
\bibitem[\protect\citeauthoryear{Cates and
  Tailleur}{2013}]{catesWhenAreActive2013}
\begin{barticle}
\bauthor{\bsnm{Cates}, \binits{M.E.}},
\bauthor{\bsnm{Tailleur}, \binits{J.}}:
\batitle{When are active {{Brownian}} particles and run-and-tumble particles
  equivalent? {{Consequences}} for motility-induced phase separation}.
\bjtitle{Europhysics Letters}
\bvolume{101}(\bissue{2}),
\bfpage{20010}
(\byear{2013})
\doiurl{10.1209/0295-5075/101/20010}
\end{barticle}
\endbibitem

%%% 46
\bibitem[\protect\citeauthoryear{Boyce
  et~al.}{2017}]{boyceElementaryDifferentialEquations2017}
\begin{bbook}
\bauthor{\bsnm{Boyce}, \binits{W.E.}},
\bauthor{\bsnm{DiPrima}, \binits{R.C.}},
\bauthor{\bsnm{Meade}, \binits{D.B.}}:
\bbtitle{Elementary {{Differential Equations}} and {{Boundary Value
  Problems}}}.
\bpublisher{{John Wiley \& Sons}},
\blocation{{Singapore}}
(\byear{2017})
\end{bbook}
\endbibitem

%%% 47
\bibitem[\protect\citeauthoryear{{Todd
  Kapitula}}{2013}]{toddkapitulaSpectralDynamicalStability2013}
\begin{bbook}
\bauthor{\bsnm{{Todd Kapitula}}}:
\bbtitle{Spectral and Dynamical Stability of Nonlinear Waves}.
\bsertitle{Applied Mathematical Sciences ({{Springer-Verlag New York Inc}}.)},
vol. \bseriesno{185}.
\bpublisher{{Springer, 2013}},
\blocation{{New York, NY}}
(\byear{2013})
\end{bbook}
\endbibitem

%%% 48
\bibitem[\protect\citeauthoryear{Burger
  et~al.}{2010}]{burgerNonlinearCrossDiffusionSize2010}
\begin{barticle}
\bauthor{\bsnm{Burger}, \binits{M.}},
\bauthor{\bsnm{Di~Francesco}, \binits{M.}},
\bauthor{\bsnm{Pietschmann}, \binits{J.-F.}},
\bauthor{\bsnm{Schlake}, \binits{B.}}:
\batitle{Nonlinear {{Cross-Diffusion}} with {{Size Exclusion}}}.
\bjtitle{SIAM Journal on Mathematical Analysis}
\bvolume{42}(\bissue{6}),
\bfpage{2842}--\blpage{2871}
(\byear{2010})
\doiurl{10.1137/100783674}
\end{barticle}
\endbibitem

%%% 49
\bibitem[\protect\citeauthoryear{Bovier and {den
  Hollander}}{2015}]{bovierMetastabilityPotentialTheoreticApproach2015}
\begin{bbook}
\bauthor{\bsnm{Bovier}, \binits{A.}},
\bauthor{\bsnm{{den Hollander}}, \binits{F.}}:
\bbtitle{Metastability: {{A Potential-Theoretic Approach}}}.
\bsertitle{Grundlehren Der Mathematischen {{Wissenschaften}}},
vol. \bseriesno{351}.
\bpublisher{{Springer International Publishing}},
\blocation{{Cham}}
(\byear{2015}).
\doiurl{10.1007/978-3-319-24777-9}
\end{bbook}
\endbibitem

%%% 50
\bibitem[\protect\citeauthoryear{Quastel
  et~al.}{1999}]{quastelLargeDeviationsSymmetric1999}
\begin{barticle}
\bauthor{\bsnm{Quastel}, \binits{J.}},
\bauthor{\bsnm{Rezakhanlou}, \binits{F.}},
\bauthor{\bsnm{Varadhan}, \binits{S.R.S.}}:
\batitle{Large deviations for the symmetric simple exclusion process in
  dimensions d{$\geq$} 3}.
\bjtitle{Probability Theory and Related Fields}
\bvolume{113}(\bissue{1}),
\bfpage{1}--\blpage{84}
(\byear{1999})
\doiurl{10.1007/s004400050202}
\end{barticle}
\endbibitem

%%% 51
\bibitem[\protect\citeauthoryear{Agranov
  et~al.}{2022}]{agranovEntropyProductionIts2022}
\begin{barticle}
\bauthor{\bsnm{Agranov}, \binits{T.}},
\bauthor{\bsnm{Cates}, \binits{M.E.}},
\bauthor{\bsnm{Jack}, \binits{R.L.}}:
\batitle{Entropy production and its large deviations in an active lattice gas}.
\bjtitle{Journal of Statistical Mechanics: Theory and Experiment}
\bvolume{2022}(\bissue{12}),
\bfpage{123201}
(\byear{2022})
\doiurl{10.1088/1742-5468/aca0eb}
\end{barticle}
\endbibitem

%%% 52
\bibitem[\protect\citeauthoryear{Kruk
  et~al.}{2021}]{krukFiniteVolumeMethod2021}
\begin{barticle}
\bauthor{\bsnm{Kruk}, \binits{N.}},
\bauthor{\bsnm{Carrillo}, \binits{J.A.}},
\bauthor{\bsnm{Koeppl}, \binits{H.}}:
\batitle{A finite volume method for continuum limit equations of nonlocally
  interacting active chiral particles}.
\bjtitle{Journal of Computational Physics}
\bvolume{440},
\bfpage{110275}
(\byear{2021})
\doiurl{10.1016/j.jcp.2021.110275}
\end{barticle}
\endbibitem

%%% 53
\bibitem[\protect\citeauthoryear{Erban and
  Chapman}{2020}]{erbanStochasticModellingReaction2020}
\begin{bbook}
\bauthor{\bsnm{Erban}, \binits{R.}},
\bauthor{\bsnm{Chapman}, \binits{S.J.}}:
\bbtitle{Stochastic {{Modelling}} of {{Reaction}}\textendash{{Diffusion
  Processes}}}.
\bsertitle{Cambridge {{Texts}} in {{Applied Mathematics}}}.
\bpublisher{{Cambridge University Press}},
\blocation{{Cambridge}}
(\byear{2020}).
\doiurl{10.1017/9781108628389}
\end{bbook}
\endbibitem

%%% 54
\bibitem[\protect\citeauthoryear{Yu
  et~al.}{2022}]{yuPerpendicularParallelPhase2022}
\begin{barticle}
\bauthor{\bsnm{Yu}, \binits{H.}},
\bauthor{\bsnm{Thijssen}, \binits{K.}},
\bauthor{\bsnm{Jack}, \binits{R.L.}}:
\batitle{Perpendicular and parallel phase separation in two-species driven
  diffusive lattice gases}.
\bjtitle{Physical Review E}
\bvolume{106}(\bissue{2}),
\bfpage{024129}
(\byear{2022})
\doiurl{10.1103/PhysRevE.106.024129}
\end{barticle}
\endbibitem

%%% 55
\bibitem[\protect\citeauthoryear{Korniss
  et~al.}{1997}]{kornissNonequilibriumPhaseTransitions1997}
\begin{barticle}
\bauthor{\bsnm{Korniss}, \binits{G.}},
\bauthor{\bsnm{Schmittmann}, \binits{B.}},
\bauthor{\bsnm{Zia}, \binits{R.K.P.}}:
\batitle{Nonequilibrium phase transitions in a simple three-state lattice gas}.
\bjtitle{Journal of Statistical Physics}
\bvolume{86}(\bissue{3}),
\bfpage{721}--\blpage{748}
(\byear{1997})
\doiurl{10.1007/BF02199117}
\end{barticle}
\endbibitem

%%% 56
\bibitem[\protect\citeauthoryear{Gon{\c
  c}alves}{2023}]{goncalvesHydrodynamicLimitsEmergence2023}
\begin{botherref}
\oauthor{\bsnm{Gon{\c c}alves}, \binits{P.}}:
Hydrodynamic limits: {{The}} emergence of fractional boundary conditions.
European Mathematical Society Magazine
(127),
5--14
(2023)
\end{botherref}
\endbibitem

%%% 57
\bibitem[\protect\citeauthoryear{Berendsen
  et~al.}{2017}]{berendsenCrossdiffusionModelMultiple2017}
\begin{barticle}
\bauthor{\bsnm{Berendsen}, \binits{J.}},
\bauthor{\bsnm{Burger}, \binits{M.}},
\bauthor{\bsnm{Pietschmann}, \binits{J.-F.}}:
\batitle{On a cross-diffusion model for multiple species with nonlocal
  interaction and size exclusion}.
\bjtitle{Nonlinear Analysis}
\bvolume{159},
\bfpage{10}--\blpage{39}
(\byear{2017})
\doiurl{10.1016/j.na.2017.03.010}
\end{barticle}
\endbibitem

%%% 58
\bibitem[\protect\citeauthoryear{Arita
  et~al.}{2017}]{aritaVariationalCalculationTransport2017}
\begin{barticle}
\bauthor{\bsnm{Arita}, \binits{C.}},
\bauthor{\bsnm{Krapivsky}, \binits{P.L.}},
\bauthor{\bsnm{Mallick}, \binits{K.}}:
\batitle{Variational calculation of transport coefficients in diffusive lattice
  gases}.
\bjtitle{Physical Review E}
\bvolume{95}(\bissue{3}),
\bfpage{032121}
(\byear{2017})
\doiurl{10.1103/PhysRevE.95.032121}
\end{barticle}
\endbibitem

\end{thebibliography}

\end{document}